\documentclass[times,twocolumn,final]{elsarticle}
\usepackage{medima}
\usepackage{framed,multirow}

\usepackage{amssymb}
\usepackage{latexsym}
\usepackage{graphicx}
\usepackage{comment}
\usepackage{amsmath,amssymb} % define this before the line numbering.
\usepackage{color}
\usepackage{subfigure}
\usepackage[]{animate}
\usepackage{animate}

\usepackage{soul}
\soulregister\cite7  
\soulregister\citep7  
\soulregister\citet7 
\soulregister\ref7 
\soulregister\pageref7  
\soulregister\cite7
\soulregister\citep7
\usepackage{multirow}
\usepackage{rotating}
\usepackage{wrapfig}
\usepackage{lineno}
\usepackage{epsfig}
\usepackage{nicefrac}       % compact symbols for 1/2, etc.
\usepackage{microtype} 
\usepackage{float}
\usepackage{amsmath}

\usepackage{enumerate}
\usepackage{enumitem} 

\usepackage{url}
\usepackage{xcolor}

\usepackage{hyperref}

\definecolor{newcolor}{rgb}{.8,.349,.1}

\journal{Medical Image Analysis}

\begin{document}

\verso{J. Wang and X. Liu \textit{et~al.}}

\begin{frontmatter}

\title{Automated interpretation of congenital heart disease from multi-view echocardiograms}%

\author[1]{\textbf{\underline{Jing \snm{Wang}}}\fnref{fn1}}

\author[3,6]{\textbf{\underline{Xiaofeng \snm{Liu}}}\fnref{fn1}}

\author[2]{Fangyun \snm{Wang}}

\author[2]{Lin \snm{Zheng}}

\author[1]{Fengqiao \snm{Gao}}

\author[1]{Hanwen \snm{Zhang}}

\author[2]{Xin \snm{Zhang}}

\author[4]{Wanqing \snm{Xie} }

\author[5]{Binbin \snm{Wang} }

\fntext[fn1]{Jing Wang and Xiaofeng Liu contributed equally to this work. }
\cortext[cor1]{Corresponding to Binbin Wang (Tel: +86 10 62173443), Wanqing Xie (Tel: +1-617-230-8174), and Xin Zhang (Tel:+86 10 59616161)}

\address[1]{Department of Medical Genetics and Developmental Biology, School of Basic Medical Sciences, Capital Medical University, Beijing, 10069, China}
\address[2]{Heart Center, Beijing Children's Hospital, Capital Medical University, National Center for Children’s Health, Beijing, 10045, China}
\address[3]{Department of Electrical and Computer Engineering, Carnegie Mellon University, Pittsburgh, PA, 15232 USA}
\address[4]{College of Mathematical Sciences, Harbin Engineering University, Harbin, China}
\address[5]{Center for Genetics, National Research Institute for Family Planning, Beijing, China}
\address[6]{Harvard Medical School, Harvard University, Boston, MA, 02215 USA}

\received{29 March 2020}
%\revised{5 December 2020}
\accepted{7 December 2020}
\availableonline{26 December 2020}
%\versionofrecord{5 January 2021}

%{Volume 69, April 2021, 101942}
\communicated{}

\begin{abstract}
%%%
{Congenital heart disease (CHD) is the most common birth defect and the leading cause of neonate death in China. Clinical diagnosis can be based on the selected 2D key-frames from five views. Limited by the availability of multi-view data, most methods have to rely on the insufficient single view analysis. This study proposes to automatically analyze the multi-view echocardiograms with a practical end-to-end framework. We collect the five-view echocardiograms video records of 1308 subjects (including normal controls, ventricular septal defect (VSD) patients and atrial septal defect (ASD) patients) with both disease labels and standard-view key-frame labels. Depthwise separable convolution-based multi-channel networks are adopted to largely reduce the network parameters. We also approach the imbalanced class problem by augmenting the positive training samples. Our 2D key-frame model can diagnose CHD or negative samples with an accuracy of 95.4\%, and in negative, VSD or ASD classification with an accuracy of 92.3\%. To further alleviate the work of key-frame selection in real-world implementation, we propose an adaptive soft attention scheme to directly explore the raw video data. Four kinds of neural aggregation methods are systematically investigated to fuse the information of an arbitrary number of frames in a video. Moreover, with a view detection module, the system can work without the view records. Our video-based model can diagnose with an accuracy of 93.9\% (binary classification), and 92.1\% (3-class classification) in a collected 2D video testing set, which does not need key-frame selection and view annotation in testing. The detailed ablation study and the interpretability analysis are provided. 

The presented model has high diagnostic rates for VSD and ASD that can be potentially applied to the clinical practice in the future. The short-term automated machine learning process can partially replace and promote the long-term professional training of primary doctors, improving the primary diagnosis rate of CHD in China, and laying the foundation for early diagnosis and timely treatment of children with CHD.}

%%%%
\end{abstract}


\end{frontmatter}

%\linenumbers


\section{Introduction}

{Congenital heart disease (CHD) is a cardiovascular malformation caused by abnormal development of the embryonic heart and vascular tissue \citep{van2011birth}. It is the most common birth defect in China and the leading cause of neonate and child death \citep{dai2011birth}. The incidence of CHD is about 0.5-0.7\% in China, among all subtypes, ventricular septal defect (VSD) and atrial septal defect (ASD) are the most common ones \citep{yang2009incidence,wu2014prevalence}. Early screening and accurate diagnosis of CHD are essential to reduce the risk of this disease \citep{pruetz2019delivery}, and most of the simple CHDs can be cured by timely surgery \citep{wu2006recent}.  With recent improvement of medical level in China, CHD has received increasing attention. Some pregnant women with environmental risk factors undergo fetal echocardiogram during pregnancy, which enables early diagnosis of this part of CHD \citep{zhang2015diagnostic,zhao2014pulse}.}

{Diagnosis of CHD during early childhood depends on multiview echocardiograms \citep{sun2015congenital}. Complex CHD often presents obvious echocardiogram abnormalities, which can be more easily recognized by doctors at primary hospitals, allowing a timely transfer of the patients to superior professional children's hospitals. 
However, owing to the lack of experienced cardiac sonographers, there remain a large number of children with delayed diagnosis of CHD, especially simple subtypes like ASD and VSD. 
The delayed or missed diagnosis may result in repeated pneumonia  \citep{luo2019outcomes}, the missing of the best timing of surgery, making a serious impact on children’s prognosis and future life \citep{chang2008missed}.}

{With sufficient data and the emergence of novel data driven learning algorithms, machine learning technology can make up for the missed diagnosis of CHD caused by the lack of professional doctors in echocardiogram \citep{litjens2019state}. Recent researches on automated analysis of the abnormality of the heart structure usually focus on the single-view two-dimensional photographs or dynamic images of the echocardiogram \citep{zheng2008four,pereira2017automated,maraci2018can}. However, the clinical diagnosis of CHD obtained from a single view can be hardly reliable, therefore a multi-view joint diagnosis is necessary \citep{lai2006guidelines}.}

{In this study, we collect a dataset with 1308 children multi-view echocardiogram video records and the key-frame of each video is labeled by an experienced doctor. With the help of this well-organized dataset, we propose to make automated diagnosis in an end-to-end manner. Although the convolutional neural networks (CNN) have shown tremendous success in processing two dimensional (2D) echocardiograms, the processing of multi-view ultrasound images is under-explored \citep{liu2019deep}. }

{A multi-channel CNN is constructed for the automated analysis of five views of ultrasound heart images, which summarizes the information in each view with the pointwise convolution. We explore the five views jointly to diagnose CHD. To the best of our knowledge, this is the first effort to use five 2D echocardiograms views to assist with the deep learning-based automated diagnosis of CHD. }

{Moreover, it is well known that deep learning is data starved. The limited training data cannot support the training of network with a large number of parameters, and usually results in overfitting. To alleviate this problem in our task, the Depthwise Separable Convolution (DSC) \citep{howard2017mobilenets} is adopted to largely reduce the network parameters. Other than the relatively limited training samples, the number of collected negative samples is usually larger than the number of positive (VSD and ASD) samples. We approach the imbalanced class problem by augmenting the positive training samples.}

{To alleviate the manual labeling of the key-frame, we further propose to process the raw videos without the need of key-frame labels and view class labels. We analyzed a series of aggregation framework to adaptively process the video data and achieve the comparable performance as the key-frame based version. Moreover, the key-frame labels in our dataset can also be utilized to guide the aggregation. Besides, a view detection module can be added to automatically classify the collected views and put them in order for the corresponding view-specific encoder.}

{This study aims to investigate the 2D echocardiograms analysis with five standard views. We collect the most common VSD and ASD as examples, and provide a serial of strong baseline methods. The contributions of this work can be summarized as follows:}

\begin{itemize}
    \item  {We collect the first large scale CHD dataset for five-view two-dimensional echocardiograms analysis. Both the raw video and the experienced doctor labeled key-frame versions are provided.} 
    
    \item  {A series of powerful baselines to explore both the key-frame-based and video-based multi-view diagnose. The depthwise separable convolution-based efficient multi-channel CNN architecture can utilize the limited data to achieve satisfactory multi-view diagnosis performance.}

    \item {In this paper, we propose a novel soft-attention framework to process video frames with or without the key-frame label in the training, while we do not need key-frame labeling and view labels in the testing stage. Four practical video aggregation schemes are investigated to explore the information among the variable number of frames.} 
 
    \item {We also demonstrate that the proposed attention mechanism provides fine-scale attention maps that can be visualized, with minimal computational overhead, which helps with the interpretability of predictions.}

\end{itemize}

\section{Related work}

\noindent \textbf{Deep learning for echocardiograms.} {With the rapid development of machine learning, automated machine analysis and interpretation technology is increasingly used in medicine \citep{liu2019deep}. More and more researchers are turning their attention to the data analysis and identification of medical images \citep{de2018clinically,esteva2017dermatologist,bejnordi2017diagnostic}.}

{Early attempts extract the handcrafted features from a cardiovascular image \citep{maraci2017framework} and fed them to the statistical classifier, e.g., support vector machine \citep{cortes1995support}. The feature for CHD prediction is usually describing the texture and shape characteristics \citep{criminisi2009decision,zheng2008four}. The kernel dynamic texture models have been applied to label each individual ultrasound image \citep{kwitt2013localizing}.}

{Recently, convolutional neural networks based deep learning gained enormous successes in image analysis tasks \citep{liu2020disentanglement,Liu_2019_ICCV}. Instead of designing the feature by humans and subsequently feeding it to a prediction model, deep learning proposes to simultaneously learn relevant features and the prediction model from the raw image in an end-to-end fashion \citep{lecun2015deep,che2019deep,He_2020_CVPR_Workshops,Han_2020_CVPR_Workshops,He_2020_CVPR_Workshops}.}

{Related studies have shown that using deep neural networks, machines can effectively identify abnormalities from ultrasonic views \citep{litjens2019state,liu2020unimodal,liu2019unimodal,liu2018ordinal,liu2020symmetric}. Diller et al. use a four-convolutional-layer CNN to process a single image to assess patients with a systemic right ventricle \citep{diller2019utility}.

At present, researchers focus on two-dimensional photographs or dynamic images of the echocardiogram and explore the intelligent recognition method and diagnostic performance for abnormality of the heart structure \citep{zheng2008four,pereira2017automated,maraci2018can}. 
The most closely related work is that of Zhang et al. \citep{zhang2018fully}, which use two neural network branches to diagnose PSLAX, PSSAX, apical 2-chamber, apical 3-chamber, and A4C images for hypertrophic cardiomyopathy and cardiac amyloid.} 

\noindent \textbf{Image set based recognition.} A video can be considered as an image set with ordered images and has been actively studied. \citep{yang2017neural,liu2017quality} compute a score for each image with neural image assessment modules. Then, a set of features are aggregated to a fixed size feature vector via weighted average pooling. Without inner-set interactions, this may result in redundancy and sacrifice the diversity in a set. \citep{liu2018dependency} proposes to exploit the inner-set relationship using reinforcement learning. However, its decision is based on the incomplete observation of the averaged features. Besides, the reinforcement learning itself is usually unstable \citep{henderson2017deep}, the feature dimension is necessary to be compacted from 1024 in its GoogleNet backbone \citep{szegedy2016rethinking} to 128 which undoubtedly weakened its representation ability.

\begin{table}[t]  \caption{The statistics of collected patient samples. VSD: ventricular septal defect;
ASD: atrial septal defect. Testing set is consisted with 82 healthy control subjects, 21 ASD subjects and 28 VSD subjects.}\label{tab:1}
\resizebox{0.9\linewidth}{!}{
\centering
\renewcommand\arraystretch{1.2}

\begin{tabular}{l|c|c|c}
\hline\hline
&Healthy control&	ASD&VSD\\\hline
A4C	& 823	&209&	276\\
PSLAX of left ventricle	&820&	171&	218\\
PSSAX of aorta&	764	&166&	243\\
SXLAX of two atria&	625&	102	&100\\
SSLAX of aortic arch&	641	&84&	99\\\hline\hline
\end{tabular}}
\end{table}

\noindent \textbf{Self-attention and non-local.}
As attention models grew in popularity, \citep{vaswani2017attention} developed a self-attention mechanism for machine translation. It calculated the response at one position as a weighted sum of all positions in the sentences. A similar idea is also inherited in the non-local algorithm \citep{buades2005non}, which is a classical image denoising technique. The interaction networks also developed for modeling the pair-wise interactions \citep{battaglia2016interaction,hoshen2017vain,watters2017visual,yang2018learning}. Moreover, \citep{wang2018non} proposes to bridge self-attention to the more general class of non-local filtering operations. \citep{zhou2017temporal} proposes to learn temporal dependencies between video frames at multiple time scales. Inspired by above works, we further adapted this idea to the multi-view echocardiograms video analysis with variable lengths.

\noindent{\bf{Temporal modeling}} is widely used in video classification, video-based identification and temporal action detection etc. \citep{zhou2017see} proposes to model the temporal information between frames using a recursive neural network (RNN). The 3D CNN is later developed to extract spatial-temporal features from video clips directly \citep{tran2015learning}. However, it is not compatible with our attention-based module. Two types of temporal attention methods have been recently developed. The first method uses spatial convolution first followed by the fully connected (FC) layers \citep{liu2017quality}. The FC layers limit the model to fixed length video clip. The second approach chooses the temporal convolution instead of FC \citep{gao2018revisiting}. \citep{gao2018revisiting} uses temporal convolution for person re-identification with fixed length clip, we show that it is promising for echocardiograms video analysis with variable lengths.

\begin{figure}[t]
\begin{center}
\includegraphics[width=1\linewidth]{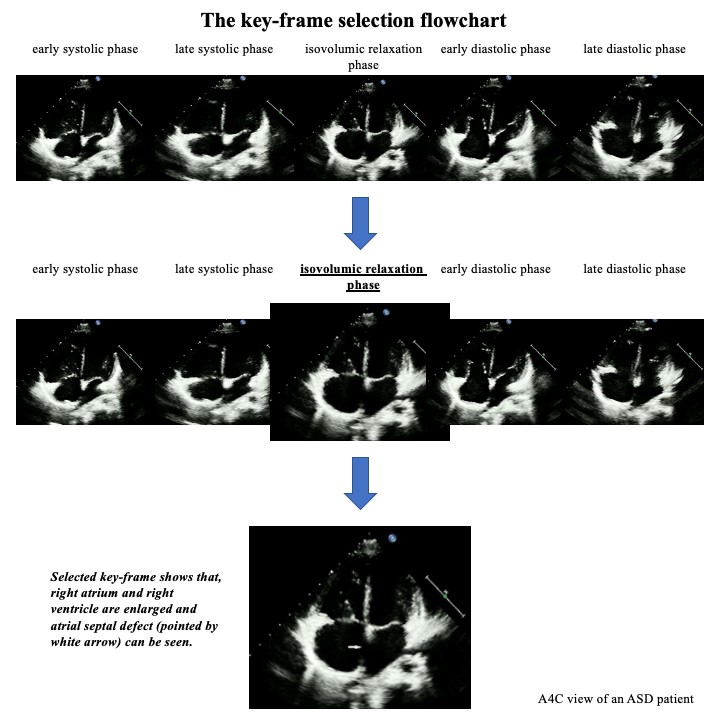}
\end{center} 
\caption{The key frame selection flowchart. We take an A4C view of an ASD patient as an example.}
\label{fig:flowchart}
\end{figure}

\section{Methodology}

\subsection{Standard collection of cardiac ultrasound image data}

Totally 1308 children (823 healthy controls, 209 VSD and 276 ASD) are collected from Beijing Children’s hospital. Each patient has the echocardiogram video records from 1 to 5 views that can be sufficient for diagnosis (shown in Table \ref{tab:1}). Besides, a single high-quality 2D frame in a video of each view is selected by the doctor to construct a 2D echocardiogram dataset.

The patient was placed in the supine position and the chest was exposed for the echocardiogram. We used PHILIPS iE 33, iE Elite, and EPIQ 7C (Philips Electronics Nederland B.V.) as instruments. The transducers frequency was ranging from 3-8 MHz. According to the heart segmental approach, the heart position, atrial position, and ventricular position were determined, and the connection relationship between the atria, ventricle, and aorta was analyzed. The atrial septum and ventricular septum were observed for whether had defects, and cavity or pulmonary venous return was noted. Five standard 2D views, the parasternal long axis view (PSLAX) of left ventricle, the parasternal short-axis view (PSSAX) of aorta, the apical four chambers view (A4C), the subxiphoid long axis view (SXLAX) of two atria and the suprasternal long-axis view (SSLAX) of aortic arch were collected. 

\begin{figure}[t]
\begin{center}
\includegraphics[width=1\linewidth]{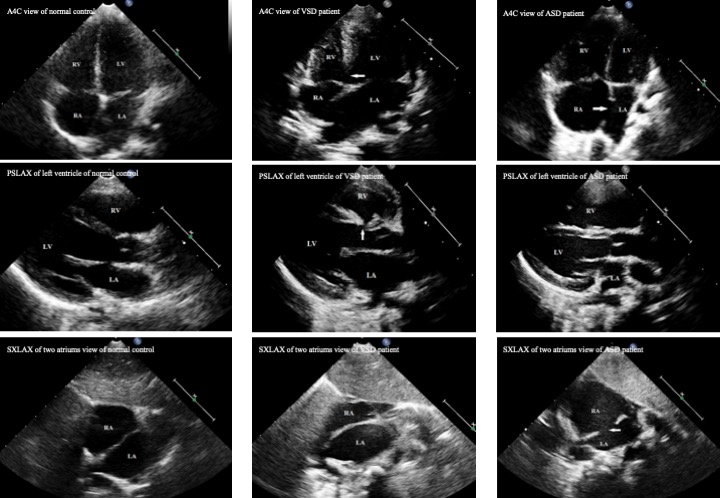}
\end{center} 
\caption{The key frame selection results of different views. From left to right: A4C, PSLAX and SXLAX view. From top to bottom: normal controls, VSD patients and ASD patients.}
\label{fig:sample}
\end{figure}

All patients’ diagnosis was confirmed by either at least two senior ultrasound doctors or intraoperative final diagnosis. The study protocol was approved by the Ethics Committee of Beijing Children’s Hospital (No. 2019-k-342).

The originally collected videos of each view are three cardiac cycles. To facilitate the processing, we also sampled a subset by randomly crop a clip of 0.8 seconds for each view.  Since the cardiac cycle of the child is typically from 0.5-0.6s, the clip of 0.8 can usually have and only have one complete cardiac cycle and one key-frame. The originally collected video dataset has a different video length for different patients and different views.

Actually, the labeling of cardiac cycles is very costly in real-world implementations. Besides, taking many cardiac cycles as input will be very redundant. Considering we are using the same instrument for collection, the temporal resolution is consistent for all of the samples. The final dataset has two versions, i.e., the original video and the cropped video. We utilize the cropped version for our video-based experiments, while the original version is possible to support more sophisticated settings.

The key-frame is manually selected by the experienced doctor. We selected the isovolumic relaxation phase as a key-frame when the ventricles finish contracting and start to relax, the defects of VSD and ASD both could be shown clearly at that time. The flowchart in Fig. \ref{fig:flowchart} could show the key-frame selection process, and the selected samples of different views are shown in Fig. \ref{fig:sample}.

We select 10\% patients to construct our fixed testing set (i.e., 82 healthy control, 21 ASD and 28 VSD patients) in a patient-independent manner. That means the videos or keyframes used in training will not be incorporated in testing. Noticing that all our selected testing patients have all the five views original video records or selected key-frame.

\begin{figure}[t]
\begin{center}
\includegraphics[width=1\linewidth]{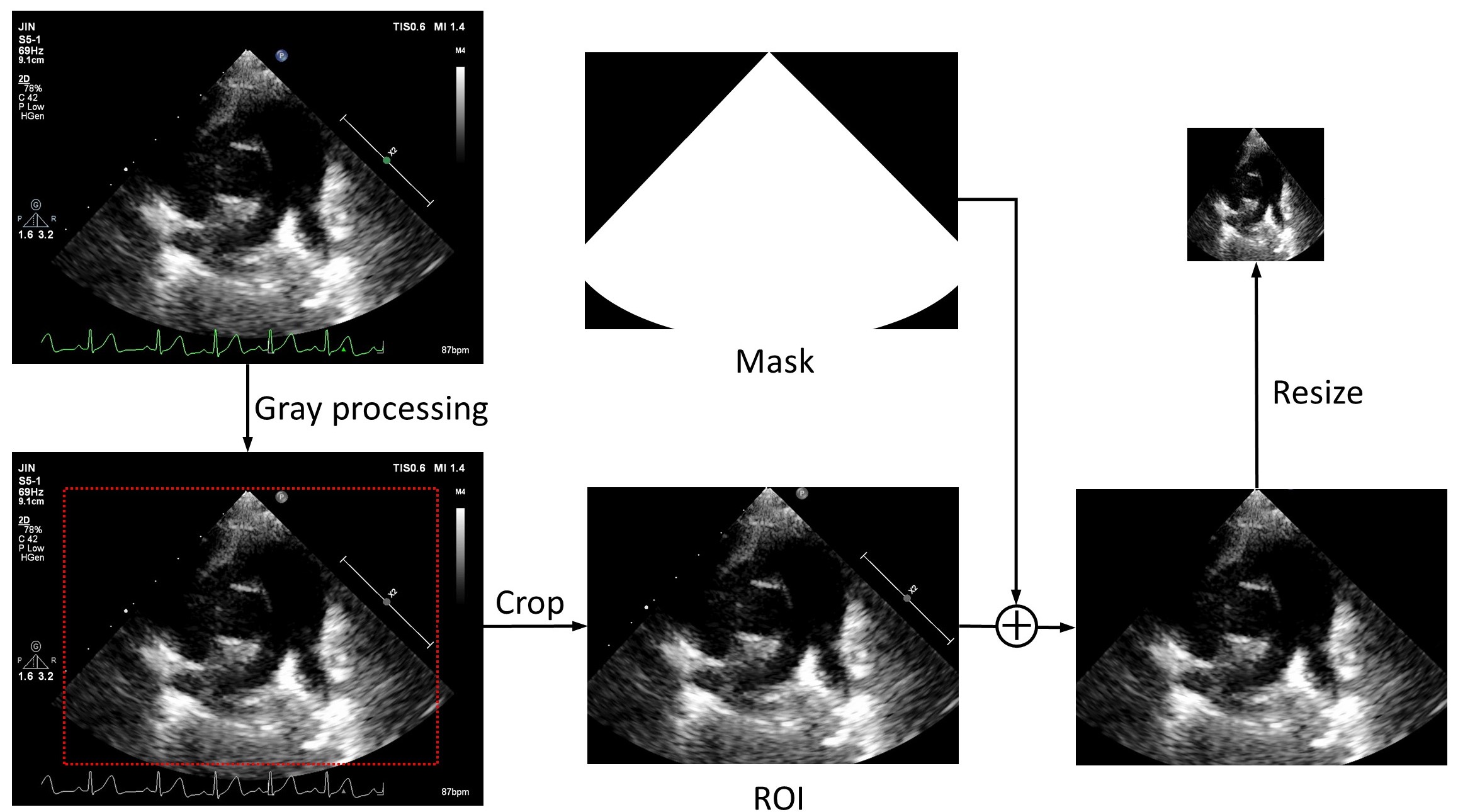}
\end{center} 
\caption{Data pre-processing flowchart.}
\label{fig:2}
\end{figure}

\subsection{Pre-processing}

Each frame in the video or the selected keyframe is a two-dimensional echocardiogram with color instrument marks, which is essentially the three-channel RGB image. We first processed each frame to a single-channel grayscale image, and the region of interest (ROI) is cropped to remove the irrelevant parts. Because the collected ultrasonic image is a circular sector and some labels cannot be removed by rectangular cropping, a mask is designed to cover these factors. This is necessary, since only the positive samples have electrocardiogram record in our dataset and the CNNs can easily discriminate the samples according to this clue. To align with the input of CNNs, the masked ROI is resized to 128$\times$128. The flow chart of pre-processing is shown in Figure \ref{fig:2}. We use the same pre-processing for the image of each view and concatenate these five images following fixed order: PSLAX of left ventricle, PSSAX of aorta, A4C, SXLAX of two atria and SSLAX of aortic arch. Noticing that each view only has one image in our 2D echocardiogram dataset.

\subsection{Key-frame based multi-view echocardiograms analysis}

\noindent\textbf{DSC for multi-view echocardiograms}

It is well known that neural networks are trained to approximate a mapping function by concatenation of multiple-layer simple non-linear functions and that the deeper structure is usually more powerful with respect to representational ability. However, deep learning is data starved. Thus, the limited training data cannot support tuning of the network with a large number of parameters, and usually results in overfitting. To alleviate this problem in our congenital cardiovascular disease diagnosis task, the Depthwise Separable Convolution (DSC) \citep{howard2017mobilenets} is adopted to largely reduce the network parameters. 

\begin{figure}[t]
\begin{center}
\includegraphics[width=1\linewidth]{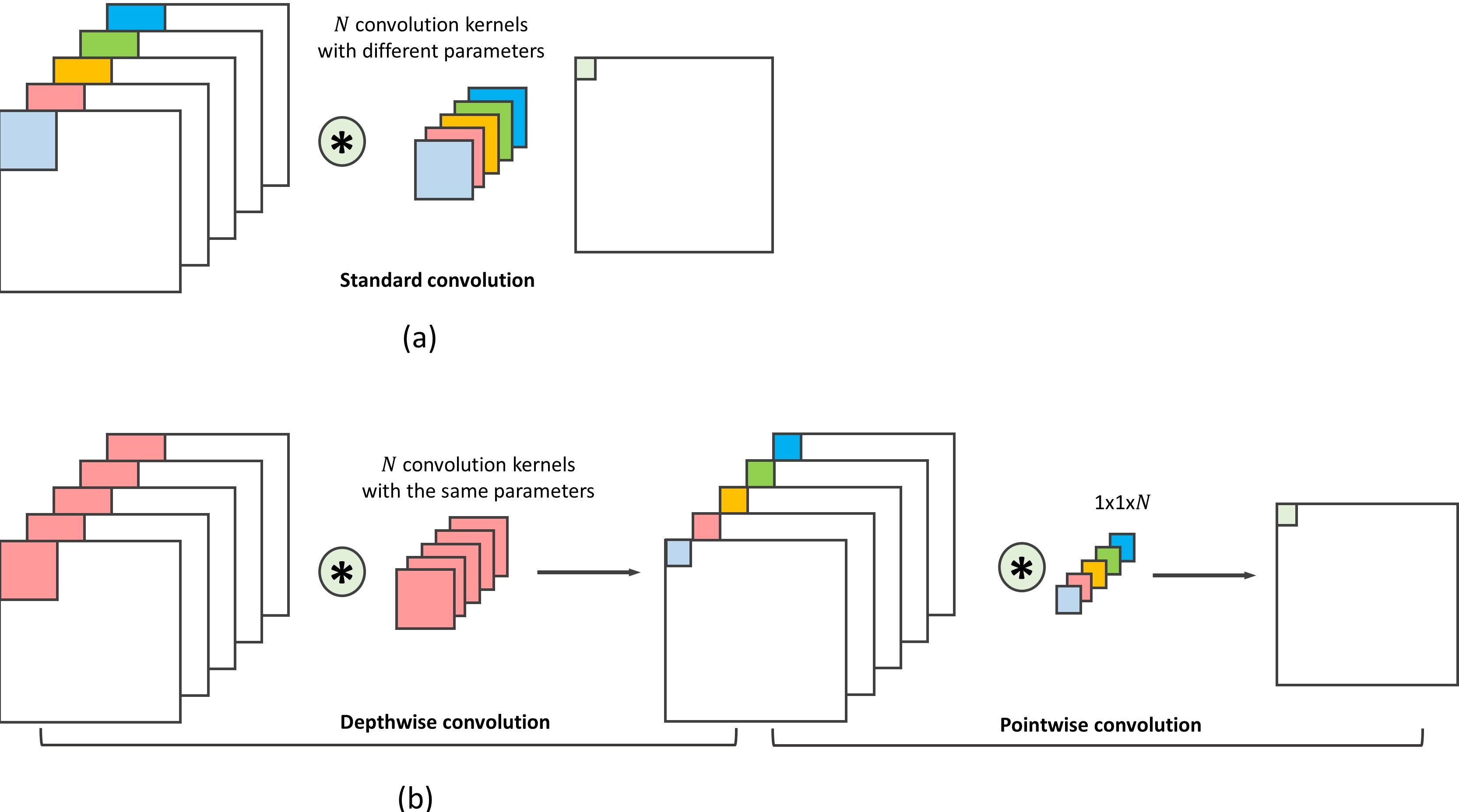}
\end{center} 
\caption{Illustration of the convolution operation. a. the standard convolution; b. the Depthwise Separable Convolution.}
\label{fig:3}
\end{figure}

We give a detailed comparison of the standard and depthwise separable convolution in Figure \ref{fig:3}. In DSC the convolution process is broken down into two operations: depth-wise convolutions and point-wise convolutions. In depth-wise operation, convolution is applied to a single channel at a time unlike standard CNN’s in which it is done for all the $N$ channels.

Noting that the DSC is originally designed for RGB data, which differs from our input, we inherit the idea of the DSC to construct our multi-channel CNN.

The widely used AlexNet \citep{simonyan2014very} has 60 million to-be-learned parameters, while the DSC (with width multiplier 0.50) \citep{howard2017mobilenets} can achieve comparable performance in the ImageNet object classification dataset using only 1.32 million parameters.

\begin{table}[t]
\caption{{Multi-channel DSC network architecture. dw denotes the depth-wise convolution with size $H\times W$, and keep the same for $N$ channels. pw denotes the point-wise convolution with size $1\times 1\times N$. We note that the first layer uses the conventional convolution with stride size 2 as in \cite{howard2017mobilenets}.}} % title of Table
\centering % used for centering table
\resizebox{0.8\columnwidth}{!}{%
\begin{tabular}{l | l | l} % centered columns (4 columns)
\hline\hline %inserts double horizontal lines
Input Size&Type / Stride & Filter Shape   \\ [0.5ex] % inserts table
%heading
\hline % inserts single horizontal line

$128\times128\times5$&Conv / s2 & 32 kernels of $3\times3\times5$  \\
\hline

$64\times64\times32$&Conv dw / s1& 32 kernels of $3\times3$ dw \\

$64\times64\times32$& Conv pw / s1& 64 kernels of$1\times1\times32$ pw \\
\hline

$64\times64\times64$& Conv dw / s2& 64 kernels of$3\times3$ dw \\

$32\times32\times64$& Conv pw / s1& 128 kernels of$1\times1\times64$ pw \\
\hline

$32\times32\times128$& Conv dw / s2& 128 kernels of $3\times3$ dw  \\

$16\times16\times128$&Conv pw / s1&128 kernels of $1\times1\times128$ pw  \\
\hline

$16\times16\times128$&Conv dw / s2&128 kernels of $3\times3$ dw \\

$8\times8\times128$& Conv pw / s1&128 kernels of $1\times1\times128$ pw \\
\hline

$8\times8\times128$ & Flatten & N/A\\

8192& FC1 & 1024 \\

1024& FC2 & 128 \\
\hline

128& Classifier & Softmax; 2 or 3-dim \\
\hline
 
\end{tabular}
\label{table:mobilenet} % is used to refer this table in the text
}
\end{table}

\noindent\textbf{Network structure design}

The diagnosis of congenital cardiovascular disease is more challenging than a common object recognition task because of its multi-view data structure as well as the relatively limited and unbalanced training sample.

\begin{table}[t]
\caption{Multi-branch DSC network architecture. Each channels are processed by the independent convolutional layers and concatenated in the first fully connected layer.} % title of Table
\centering % used for centering table
\resizebox{1\columnwidth}{!}{%
\begin{tabular}{l | l | l} % centered columns (4 columns)
\hline\hline %inserts double horizontal lines
Input Size&Type / Stride & Filter Shape   \\ [0.5ex] % inserts table
%heading
\hline % inserts single horizontal line

Five $128\times128\times1$&Conv / s2 & 5-group of 32 kernels of $3\times3\times1$  \\
\hline

Five $64\times64\times32$&Conv dw / s1& 5-group of 32 kernels of $3\times3$ dw \\

Five $64\times64\times32$& Conv pw / s1& 5-group of 64 kernels of$1\times1\times32$ pw \\
\hline

Five $64\times64\times64$& Conv dw / s2& 5-group of 64 kernels of$3\times3$ dw \\

Five $32\times32\times64$& Conv pw / s1& 5-group of 128 kernels of$1\times1\times64$ pw \\
\hline

Five $32\times32\times128$& Conv dw / s2& 5-group of 128 kernels of $3\times3$ dw  \\

Five $16\times16\times128$&Conv pw / s1& 5-group of 128 kernels of $1\times1\times128$ pw  \\
\hline

Five $16\times16\times128$&Conv dw / s2& 5-group of 128 kernels of $3\times3$ dw \\

Five $8\times8\times128$& Conv pw / s1& 5-group of 128 kernels of $1\times1\times128$ pw \\
\hline

Five $8\times8\times128$ & Flatten & N/A\\

40960 & FC1 & 5120 \\
5120& FC2 & 128 \\
\hline

128& Classifier & Softmax; 2 or 3-dim \\
\hline
 
\end{tabular}
\label{table:mobilenet} % is used to refer this table in the text
}
\end{table}

\begin{figure}[t]
\begin{center}
\includegraphics[width=1\linewidth]{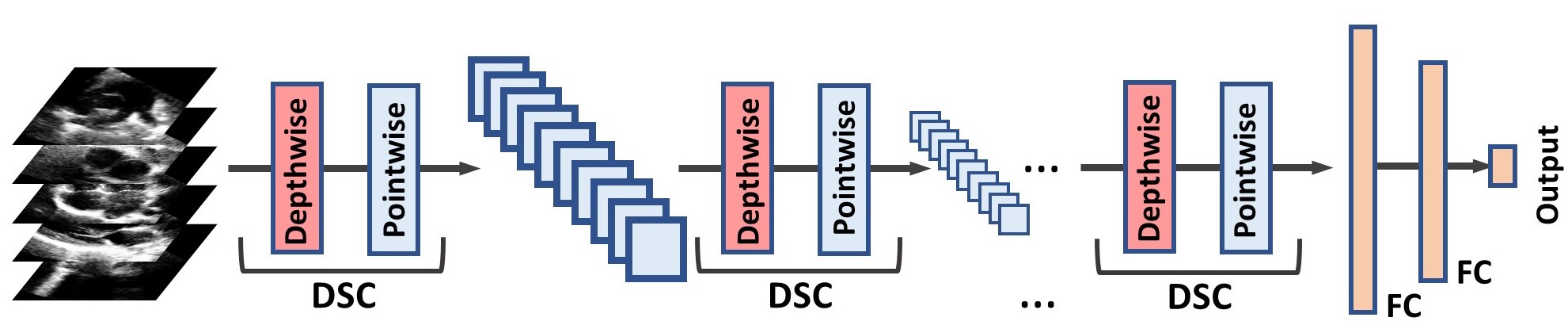}
\end{center} 
\caption{Illustration of the multichannel convolutional neural networks for key frame-based multi-view diagnosis.}
\label{fig:4}
\end{figure}

The decision is based on five views, which can incorporate more complementary information than a single view-based diagnosis. However, this also introduces the challenge of information fusion. We propose to concatenate the five views sequentially and produce a matrix with size 128$\times$128$\times$5. A five-channel CNN, as shown in Figure \ref{fig:4}, is developed to take the concatenated matrix as input. Each DSC block incorporates a depthwise convolution and a pointwise convolution. After a few DSC layers, the feature maps will be flattened as the feature vector that will be processed by two fully connected layers with sizes 1024 and 128, respectively. The detailed network structure is given in Table 2.

We note that the multi-branch framework can be an alternative to the multi-channel network. In the multi-branch framework \citep{lee2016multi}, each of the five views is passed through an independent neural network to create high-level features, which are then combined at the end of the network before making a final prediction. 

Following the comparison setting in \citep{lee2016multi}, we configure the convolutional layers for the view-independent branch, and the fully connected layers for the merged feature processing network. Considering the input sample of each branch has a size of $128\times128\times1$, we change the first convolutional layer to the 32 kernels with the size of $3\times3\times1$ kernels. With five branches, we have five $8\times8\times128$ to be fused feature maps. As in \citep{lee2016multi}, we concatenate the five flatted 8192-dimensional vectors and followed by the first fully connected layer. The detailed network structure of multi-branch DSC is detailed in Table 3.

{Our paper proposes to explore the CHD diagnosis with multi-view echocardiograms, and develop a powerful baseline model. The multi-channel scheme can adaptively learn the information fusion in each layer, instead of the simple concatenation in a fully connected layer \citep{lee2016multi}. Actually, the different views of echocardiograms also share some similarities. Therefore, the filters learned in a view may potentially be helpful in the other view. Moreover, the multi-channel model with a single-branch can have significantly fewer parameters than the multi-branch counterpart. The fewer parameter is important for the limited training data. Besides, the memory cost of the multi-channel module is much fewer than the multi-branch module. We note that the multi-channel model is not applicable for heterogeneous inputs (e.g., image and EEG data), since they require different network structure for different inputs. With sufficiently training data, the performance of multi-branch and multi-channel are usually similar for homogeneous input \citep{lee2016multi}.}

For binary classification (negative or positive), we use the sigmoid function as the output unit and optimize the network with binary cross-entropy loss. 

For the three-class classification (negative, ASD and VSD), we use the transfer learning from binary task. We choose the best pretrained binary classification model and configure its output layer as three neural units followed by a SoftMax function. The network is fine-tuned with multiclass cross-entropy loss. To save time, the CNNs could initially be trained to solve the 3-class task and their outputs could be then merged to get the binary prediction results.

\noindent\textbf{Balancing classes with data augmentation}
In the three-class classification case, the number of collected negative samples is almost four times larger than that of VSD or ASD. The richer source of the healthy control sample is common in many medical tasks. However, the unbalanced data distribution potentially results in the inference of network bias to the negative class. Therefore, we propose augmentation of the positive training samples by constructing “virtual” positive patients using a different combination of the original positive patients in the training set. We note that the testing set is always fixed with the original five views. Specifically, for the training samples without all five views, we randomly sample the missed views from the other patients in the same class (i.e., negative, ASD and VSD) to construct more patient samples. Since the original 1 to 4 views can be sufficient for diagnosis, the added view from the same class will not change the disease class. In our experimental settings, we construct 4 times positive (VSD+ASD) samples of the original positive patients to achieve a balanced training distribution.

\subsection{Video based multi-view echocardiogram analysis}

Keyframe selecting can always be costly, and the automatic analysis of multi-view videos in an end-to-end fashion is necessary when large scale samples are collected. Therefore, we propose to investigate the possible deep aggregation frameworks for our video-based multi-view two-dimensional echocardiogram analysis. The basic idea is to assign a larger weight for the more important frame in the diagnosis procedure.

\begin{figure*}[t]
\begin{center}
\includegraphics[width=0.8\linewidth]{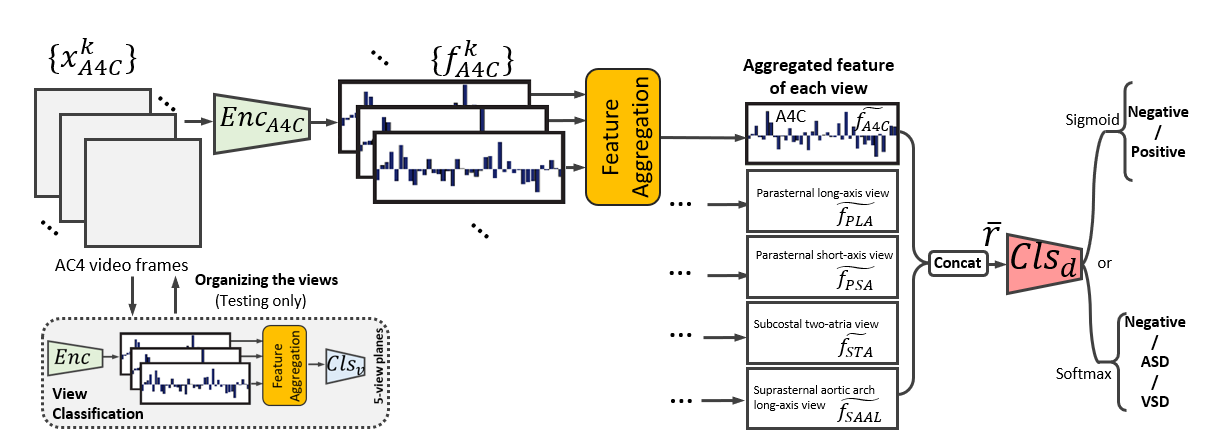}
\end{center} 
\caption{The overall framework of video-based multi-view two-dimensional echocardiograms analysis. The view classification module adopts the same backbone as the view-independent encoder.}
\label{fig:1}
\end{figure*}

The overall framework is shown in Figure \ref{fig:1}. The video of each view is feed to a view-specific encoder to extract the frame-independent feature representations. For example, the $k$ A4C frames $x_{A4C}^k$ is encoded to $k$ $f_{A4C}^k$ feature vectors. This is a typical solution for dimension reduction and information compressing \citep{liu2019dependency,liu2018dependency}. Specifically, it embeds the images into latent space independently and generates the corresponding feature sets $\{f_{A4C}^k\}$, where $f_{A4C}^k \in \mathbb{R}^{H\times W\times D}$, $k\in{1,2,\cdots K}$ indicates the $K$ frames, $H$, $W$ and $D$ are the height, width and channel dimension of our representation, respectively. Empirically, we use the convolutional layers of key-frame based model DSC as our encoder (e.g., $Enc_{A4C}$), and the size of $x_{A4C}^k$ is $H=8, W=8, D=128$.

\begin{figure}[t]
\begin{center}
\includegraphics[width=0.8\linewidth]{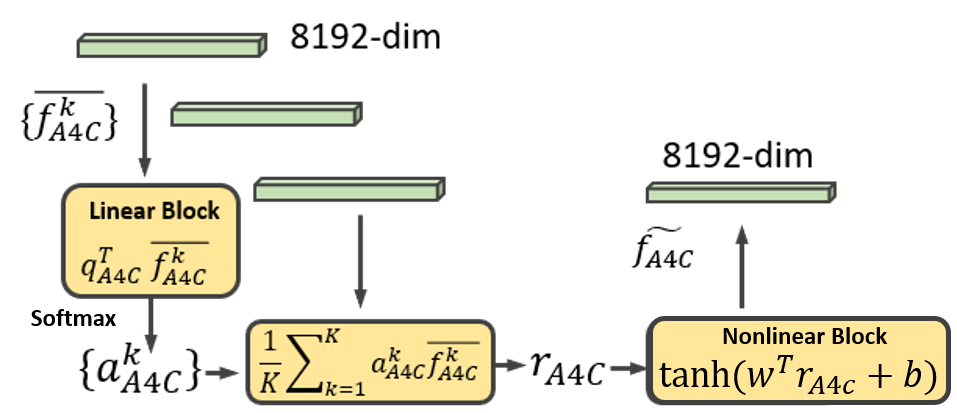}
\end{center} 
\caption{The frame-independent aggregation module used for video-based echocardiogram analysis.}
\label{fig:5}
\end{figure}

Then, a feature aggregation module is applied to the features of each frame and produces a representation of this view. The ordered representations for all five views are concatenated to the later disease classifier $Cls_d$ for binary or three-class classification. We adapted four possible feature aggregation modules for our video-based echocardiogram analysis.

\noindent\textbf{Frame-independent aggregation}

{The first is frame-independent aggregation \citep{yang2017neural}, which uses a neural quality assessment module to assign a weight for each frame (Figure \ref{fig:7}). Then, the features are averaged according to the weight. The $f_{A4C}^k$ is flattened to an 8192-dimensional vector $\overline{f_{A4C}^k}$ and followed by the weight assignment network \citep{yang2017neural}.} We also chose the two-block neural network as in \citep{yang2017neural} for weight assignment.

The first attention block is the linear transformation with softmax normalization $a_{A4C}^k=\frac{{\text{exp}(e_{A4C}^k)}}{\sum_j{\text{exp}(e_{A4C}^j)}}$ and $e_{A4C}^k={q_{A4C}}^T \overline{f_{A4C}^k}$, where $a_{A4C}^k\in\mathbb{R}$, vector $q_{A4C}$ has the same size as a single feature $\overline{f_{A4C}^k}$. Then, the weighted sum of $\overline{f_{A4C}^k}$ with the weights $a_{A4C}^k$ is calculated to form the aggregated feature $r_{A4C}=\frac{1}{K}\sum_k a_{A4C}^k \overline{f_{A4C}^k}$, where $r_{A4C}$ is also a 8192-dimensional vector, $k$ is the index of $K$ frames.

{With the linear aggregated $r_{A4C}$, the nonlinear block applies the tanh operation to calculate the final aggregation $\widetilde{f_{A4C}}=\text{tanh}(w^T r_{A4C}+b)$, where $w$ and $b$ are the weight and bias term of linear neural unit.}

{The aggregated 8192-dimensional feature representation $\widetilde{f_{A4C}}$ is concatenated with the feature vector of the other views and followed by the fully connected layer. We also use the two-layer fully connected layers with the size of 1024 and 128 as in image-based classification settings. The frame-independent weighting scheme has a simple structure and each frame can be processed parallel. However, the assessment is only based on a single frame input and cannot consider the inter-frame relationship. Actually, the neighboring frames in the echocardiogram video are closely related and follows the cardiac cycle.}

\noindent\textbf{Recurrent neural network}

{The recurrent neural network (RNN) is a well-established method for video processing. Although its training can be unstable and the processing is relative slow \citep{yang2017neural,liu2018dependency}.}

\begin{figure}[t]
\begin{center}
\includegraphics[width=0.8\linewidth]{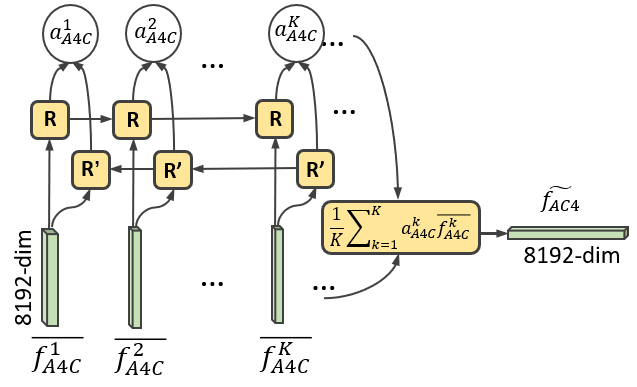}
\end{center} 
\caption{The recurrent neural network module used for video-based echocardiogram analysis.}
\label{fig:6}
\end{figure}

\begin{figure}[t]
\begin{center}
\includegraphics[width=1\linewidth]{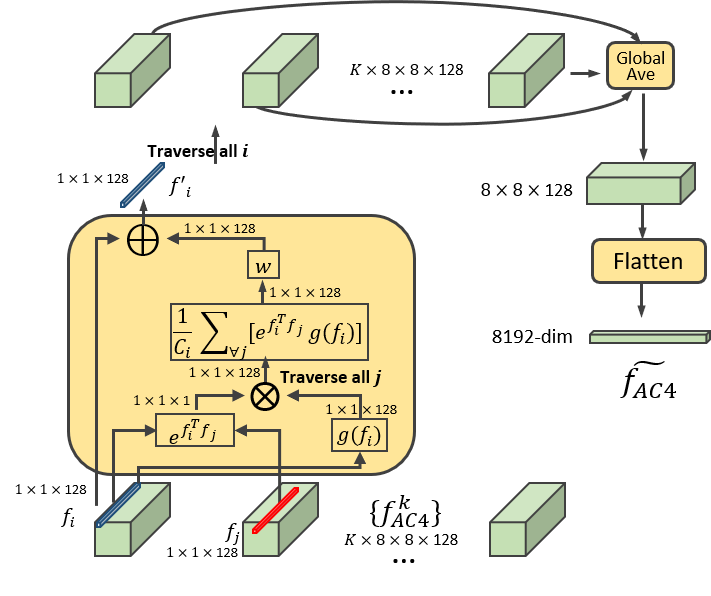}
\end{center} 
\caption{The non-local aggregation module used for video-based echocardiogram analysis.}
\label{fig:7}
\end{figure}

{We propose to use RNN to assign the weight of each frame by considering the correlations with the neighboring frames. We adapt our video-based attention scheme based on RNN as shown in Figure \ref{fig:6}. The $f_{A4C}^k$ is also flattened to an 8192-dimensional vector $\overline{f_{A4C}^k}$ and followed by the RNN.} 

Specifically, the bi-directional long short-term memory network is employed as the recurrent layer, which takes the sequential feature vectors as inputs and produces a sequence of activations. Then, the activations are normalized via the softmax. It is a typical solution for sequential data, but usually takes much more processing time in both training and testing, and hard to train \citep{liu2019permutation}. Actually, the cardiac cycle in echocardiogram videos is important for keyframe selection.

\noindent\textbf{Non-local aggregation}

{Considering the training difficulty of RNN and information loss along with the recurrent scheme, we propose to utilize the non-local aggregation framework which can take an arbitrary number of frames as input and adaptively learn a complementary representation by considering the redundancy between each frame. A shown in Fig. \ref{fig:7},  the non-local block receives $K$ extracted features ${\{f_{A4C}^k\}}$ and restructuring them based on inner-set correlations. ${\{f_{A4C}^k\}}$ are deterministically computed from the image set ${\{x_{A4C}^k\}}$, they also inherit and display large variations and redundancy.}

{For $K$ feature maps, there are ${\small K}\times {\small 8}\times {\small 8}$ positions in the HW plane. Following \citep{wang2018non}, we use $i\in\{1,2,\cdots,{\small K}\times {\small 8}\times {\small 8}\}$ to index the position whose response is to be computed, and $j$ is the index that enumerates all of the other possible positions.}

{We omit the superscript $k$ and subscript $A4C$ for simply and use $f_i\in\mathbb{R}^{1\times1\times128}$ denote the feature vector at position $i$. After the to be processed $f_i$ is chosen, there are $K\times8\times8-1$ possible to be compared features which are denoted by $f_j\in\mathbb{R}^{1\times1\times128}$. The non-local block can be formulated as}

%Besides, for each $i$, we use $j$ to index all $D$-dimensional feature vectors other than the $i^{th}$ vector $j\in\{1,2,\cdots,{{\small K}\times \small H}\times {\small W}-1\}$. 

\begin{equation}
f_{i}'=w[\frac{1}{C_{i}}\sum_{\forall {j}}e^{f_i^Tf_j}g(x_i)]+f_{i}\nonumber\label{con:2}\vspace{-10pt}\end{equation}\begin{align}{C_{i}}=\sum_{\forall j}e^{f_i^Tf_j}
\end{align}

{\noindent where the scalar $e^{f_i^Tf_j}\in\mathbb{R}$ is the dot-product similarity of $f_i$ and $f_j$. We can also use the Euclidean distance to model the relationship of $f_i$ and $f_j$, but the dot-product similarity is more implementation-friendly in modern deep learning platforms. $g(f_i)$ is an embedding function, which applies the point-wise $1\times 1\times 128$ convolutions to $f_j$, and output a vector with the size of $1\times 1\times 128$. ${C_{i}}$ is a normalization term. ${w}\in \mathbb{R}^{1\times1\times 128}$ is a to be learned weight vector.}

{For a specific $f_i$, we traverse its $K\times8\times8-1$ possible neighboring $f_i$ to compute $\frac{1}{C_{i}}\sum_{\forall {j}}e^{f_i^Tf_j}g(x_i)$. After the processing of a non-local block, $f_i$ is reconstructed to $f_i'\in\mathbb{R}^{1\times1\times128}$. By traversing all ${K\times8\times8}$ $f_i$, the non-local block output $K$ feature maps, each with the size of ${8\times8\times128}$. Then, the global pooling \citep{wang2018non} is applied for these $K$ feature maps to calculate the element-wise average.}

\begin{figure}[t]
\begin{center}
\includegraphics[width=0.9\linewidth]{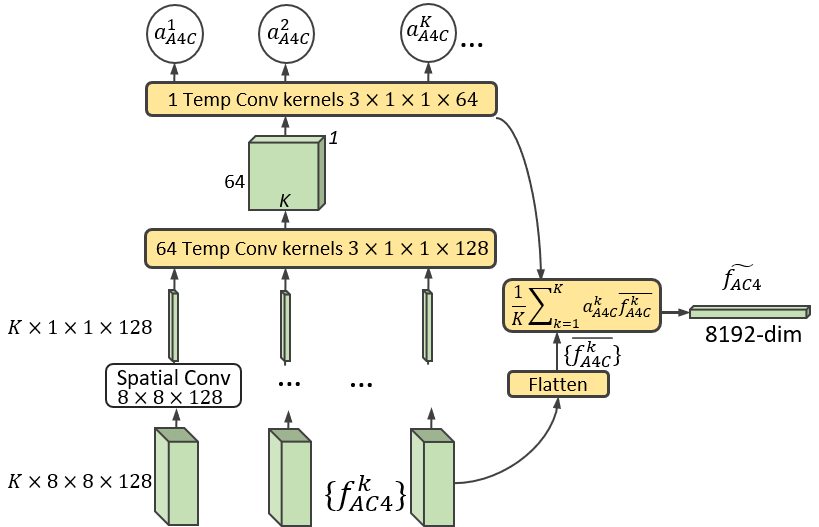}
\end{center} 
\caption{The temporal convolution module used for video-based echocardiogram analysis.}
\label{fig:9}
\end{figure}

\noindent\textbf{Temporal convolution}
{Considering the complicated processing of the RNN and non-local modules, we propose to utilize the temporal convolution \citep{bai2018empirical} to explore the neighboring relationships. Specifically, we adapt it to assign the attention score for aggregations as shown in Figure \ref{fig:9}. It is introduced as a more efficient alternative for RNN \citep{gao2018revisiting}.}

{Following \citep{gao2018revisiting}, we first apply 64 spatial convolution kernels with the size of $8\times8\times128$ for each input feature map as the first layer to summarize the spatial information and produce a $1\times1\times128$ feature vector for each frame. We note that the spatial convolution is shared for all frames.}

{Then, we concatenate $K$ feature vectors to a spatial-temporal feature map with the size of $k\times1\times1\times128$. It is followed by two temporal convolution layers with padding and stride size is set to 1. The first layer uses 64 temporal convolution kernels with the size of $3\times1\times1\times128$, where 3 is for the time sequence direction, and the last three-dimension matches the size of the input frame-wise feature vector. Actually, each temporal convolution kernel output a $K$-dimensional vector which can be directly used as the attention score. However, the convolutional window for time direction is limited to three. To further enlarge the reception field in the time direction, we adopt the multiple temporal convolution layers as in \citep{bai2018empirical}. For the input of 64-dim features, i.e., the input matrix has the size ${K\times1\times1\times64}$, the second layer temporal convolution kernel has the size of ${3\times1\times1\times64}$. The dimension of the output vector is equal to the number of frames in a video $K$. We use softmax to get the normalized attention score.}

We note that the receptive field of temporal convolution is relatively small compared to the RNN. However, the redundancy in the video is mostly confined to neighboring frames. It can be a good balance of exploring the sequential relationship and processing speed.

\noindent\textbf{Network and training details.} The encoder and disease classifier $Cls_d$ use the same backbone as the image-based setting (i.e., AlexNet or DSC) for a fair comparison. The key-frame label in the training set can also be used as an additional signal to guide the neural aggregation modules, e.g., frame-independent aggregation, recurrent neural network, temporal convolution. Actually, the key-frame-based diagnosis is essentially a hard attention scheme that assigns 100\% attention in the key-frame and 0\% for the others. Since we know which frame should have a large weight (i.e., 100\%), the difference between the predicted weights and the expert labeled weight can be used as an additional L2 loss for optimization.

{\subsection{View classification}}

{Moreover, the view-specific encoder requires the collected five videos to be ordered with their view labels. A practical way is letting the user or clinician note the view in the collection stage or label the view based on collected echocardiograms. To alleviate the labeling task in real-world application and double-check the possible mislabeling of views, we propose an additional multi-view classification module.

The view classification module is used to predict the five echocardiograms views. Based on the view classification result, the five views are ordered and feed to their corresponding view-specific encoders. 

The backbone structure of view classification module is the same as the normal/patient classifier, but without the multi-view feature concatenation and the $Cls_v$ has five output units for five views instead of 2 or 3.We note that the aggregation in view-classification is the same as the diagnosis network in all of our network settings.  

We note that the dataset has one video for each view. In some cases, more than one video may be classified to the same view and some of the other views are empty. For example, two videos’ $Cls_v$ softmax prediction has the maximum probability for the PSLAX view. Considering the softmax probability well calibrates the prediction confidence \cite{zou2019confidence}, we compare the maximum probability and choose the larger one as the PSLAX view, while another video is assigned to the empty view. 

The view classification for image-based diagnose is relatively redundant, since the user/doctor needs to know the view, and then make the key-frame selection. With the view classifier, our video-based diagnose system can take unordered five videos as input and automatically make the prediction.}

\begin{table}[t]  \caption{Performance of the disease classification task. The threshold for binary classification is 0.5, while we use $argmax$ function to choose the largest probability in softmax output as the prediction of 3-class classification. VSD: ventricular septal defect; ASD: atrial septal defect; A4C: the apical four chambers view; DSC: Depthwise Separable Convolution.}\label{tab:1}
\resizebox{1\linewidth}{!}{
\centering
\renewcommand\arraystretch{1.2}

\begin{tabular}{l|c c|c}
\hline\hline
\multirow{4}{*}{Methods} &\multicolumn{2}{c}{Binary classification}&\multicolumn{1}{c}{3-class classification}\\  

&\multicolumn{2}{c}{ (Negative / VSD+ASD) } &  \multicolumn{1}{c}{  (Negative / VSD / ASD)}\\\cline{2-4}

&ACC&AUC&ACC\\\hline

AlexNet (A4C view only) & 0.840 &0.806& 0.817\\\hline
AlexNet (Multi-branch) & 0.893 &0.843& 0.876\\\hline
AlexNet & 0.895 &0.845& 0.878\\\hline

DSC (A4C view only) & 0.885&0.813 & 0.855\\\hline
DSC (Multi-branch) & 0.946 &0.918& 0.916\\\hline
DSC ($\frac{1}{5}$Multi-branch) & 0.935 &\ 0.904& \ 0.907\\\hline
DSC & 0.947 &0.918& 0.917\\\hline
DSC (w/o transfer learning) & 0.939 &0.924 &0.908\\\hline
DSC+Augmentation & 0.954 &0.942& 0.923\\\hline\hline
\end{tabular}}

\end{table}

\begin{figure}[t]
\begin{center}
\includegraphics[width=0.6\linewidth]{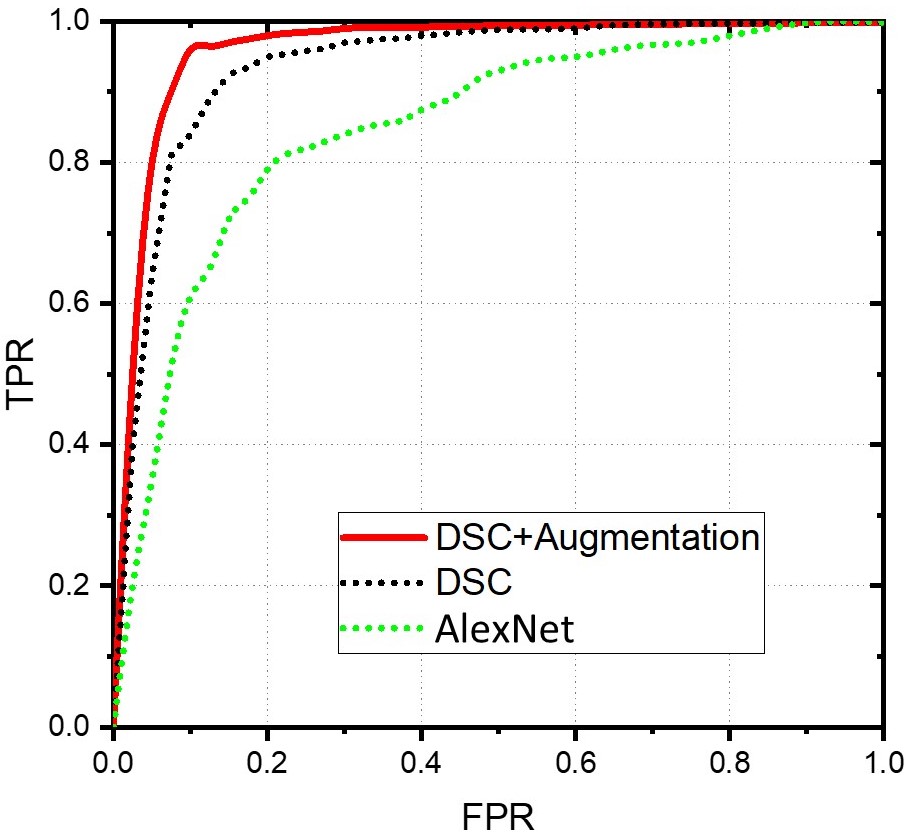}
\end{center} 
\caption{Comparison of the DSC and AlexNet baseline with respect to ROC curves in the binary classification setting, when using five convolutional layers and a 128$\times$128 input image size.}
\label{fig:10}
\end{figure}

\section{Experiments and results}

\begin{table}[t]
\begin{center}\caption{Confusion matrix of the key-frame-based binary classification
task with DSC+Augmentation when setting the threshold with respect to
\(\widetilde{y}\)= 0.5.}\label{tab:2}
\resizebox{0.6\linewidth}{!}{
\begin{tabular}{l|c|c}
\hline\hline
&$y$=Negative&$y$=Positive\\\hline
\(\widetilde{y}\)= Positive & 0.061  &   0.980\\\hline
\(\widetilde{y}\)= Negative & 0.939 & 0.020\\\hline\hline
\end{tabular}}
\end{center}
\end{table}

\subsection{2D key-frame testing set analysis}

To investigate the diagnostic performance, we evaluate the neural network system in both binary and three-class classification settings. We choose the AlexNet as our baseline, which is a widely used structure but has significantly more parameters. {As shown in Table 4, use of the DSC as the backbone outperforms the AlexNet baseline by 5.4\% with respect to testing accuracy in the binary classification setting. More appealing, DSC is 45 times smaller and 9.4 times less compute than AlexNet. The positive data augmentation can efficiently improve the performance without sophisticated algorithms. }

{We note that the results of multi-channel network is comparable or even better than the multi-branch model. Moreover, the multi-channel model has a fifth of parameters in the convolutional layers.}

Moreover, we propose to configure a multi-branch DSC network that has roughly the same parameters of our multi-channel DSC. Specifically, we use 7, 7, 14, 14, 28, 28, 28, 28, 28 kernels for each DSC layers instead of 32, 32, 64, 64, 128, 128, 128, 128, 128 respectively. Moreover, we set the first fully connected layer to the 1024-dimension. We denote this setting as DSC($\frac{1}{5}$Multi-branch) in Table 4, which is significantly worse than the multi-channel DSC. Largely reduce the parameters can significantly affect the expressivebility of the neural network and result in the performance drop of the vanilla multi-branch network.

{By pre-train the network with binary classification and then modifying the sigmoid output unit to a three-way softmax unit, we can re-train the network in the three-class classification setting. The accuracy of three-class classification is usually lower than the binary counterpart since more fine-grained discrimination is required. While the pre-training with binary classification can be helpful for the performance.} 

\begin{figure}[t]
\begin{center}
\includegraphics[width=1\linewidth]{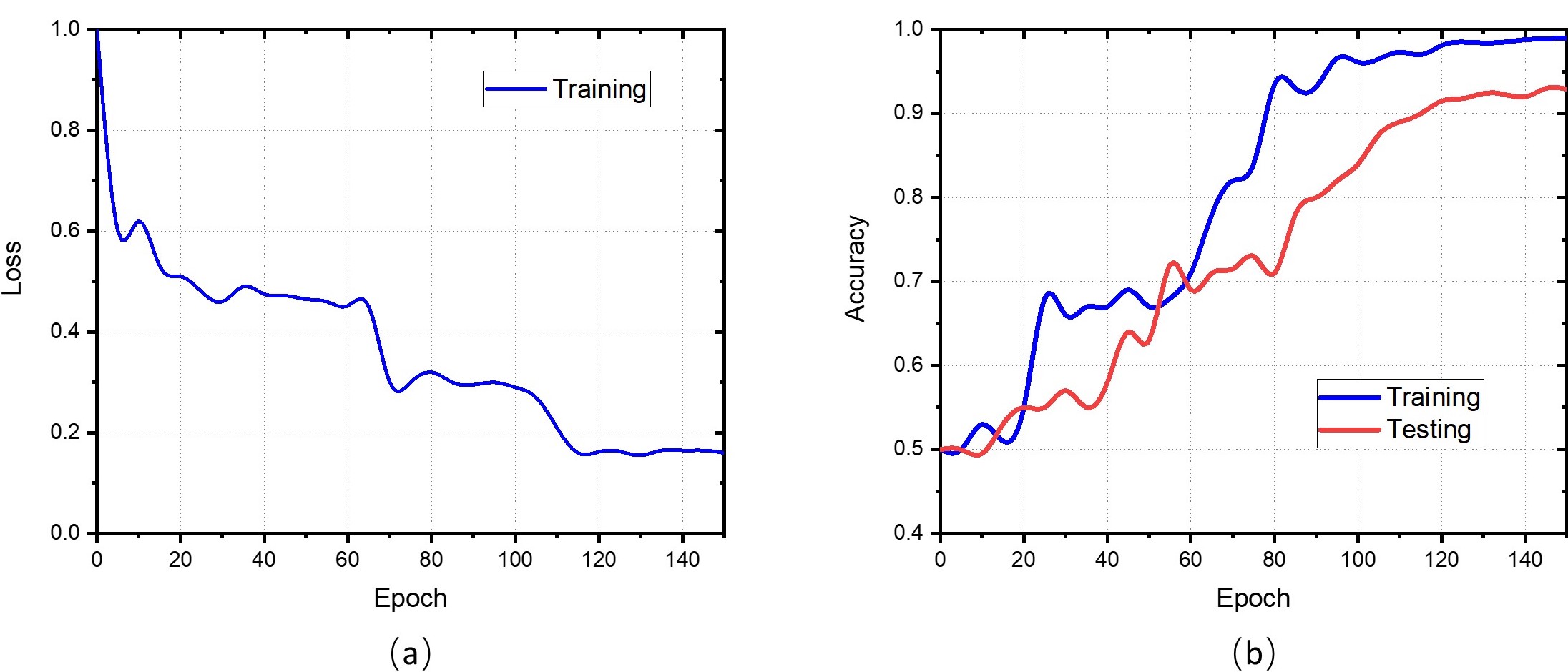}
\end{center} 
\caption{The accuracy with respect to the number of epochs for our DSC. a. training loss; b. training validation accuracies.}
\label{fig:11}
\end{figure}

To evidence the effectiveness of using five views as input rather than only using A4C view, we also propose the performance of a baseline that only use A4C view image as a single-channel input to AlexNet and our DSC networks. {The improvements of the accuracy are more than 6.2\%. We note that the proposed data augmentation does not apply to the single-channel case. In Table 4, we show the present model can diagnose positive or negative sample with an accuracy of 95.4\%, 3-class classification (negative, ASD and VSD) with an accuracy of 92.3\%. From the confusion matrix in Table 5, we can see that 98.0\% of ASD/VSD samples are correctly diagnosed as positive.}

The receiver-operating characteristic (ROC) curve plots the true positive rate (TPR) against the false positive rate (FPR) by varying a threshold. It illustrates the diagnostic ability of a binary classification system. A larger area under the curve indicates better performance. From Figure \ref{fig:10}, we can see that the DSC outperforms the AlexNet baseline by a large margin, and that the data augmentation can further improve the performance consistently.

The loss value and training validation accuracies concerning the number of epochs are shown in Figure \ref{fig:11}. The DSC converges to a steady optimum after about 120 epochs. Moreover, the training becomes saturated in the middle stage, but does not settle down at the saddle point, which can be attributed to the Adam optimizer. The DSC yields reliable validation accuracies without overfitting to the training data.

\begin{table}[t]\caption{The video-based binary and 3-class classification accuracy and testing time. \dag Using the ground truth view label directly. *The key-frame annotation in the training set is used for the aggregation module training.}\label{tab:3}
\resizebox{1\linewidth}{!}{
\centering
\renewcommand\arraystretch{1.2}

\begin{tabular}{l|c|c|c|c}
\hline\hline
&Accuracy&AUC&Accuracy& Testing time\\
&(Binary)&(Binary)&(3-class)& (Titan X)\\\hline

Frame-independent & 0.908 & 0.894 & 0.893 & 98ms\\
Frame-independent\dag & 0.913 & 0.898& 0.901 & 42ms\\
Frame-independent* & 0.915 & 0.902& 0.903 & 98ms\\\hline

RNN & 0.918& 0.909 & 0.912 & 1537ms\\
RNN\dag & 0.921 & 0.912 & 0.916 & 716ms\\
RNN* & 0.931 & 0.917& 0.918  & 1537ms\\\hline

Non-local  & 0.933 &0.918 & 0.914 & 294ms\\
Non-local\dag  & 0.935 &0.920 & 0.918 & 187ms\\\hline

Temporal convolution & 0.939 & 0.922 & 0.921 & 132ms\\
Temporal convolution\dag & 0.944& 0.926 & 0.923 & 60ms\\
Temporal convolution* & 0.947& 0.928 & 0.925 & 132ms\\\hline\hline
\end{tabular}}
\end{table}

\begin{table}[t] \caption{Confusion matrix of the video-based binary classification task with DSC+Augmentation (without key-frame label for training) when setting the threshold to \(\widetilde{y}\) = 0.5.}\label{tab:4} 
\resizebox{0.9\linewidth}{!}{
\centering
\renewcommand\arraystretch{1.2}

\begin{tabular}{l|c|c}
\hline\hline

Frame-independent aggregation & \(y =\) Negative & \(y =\) Positive\\\hline
\(\widetilde{y} =\)Positive & 0.073 & 0.918\\
\(\widetilde{y} =\)Negative & 0.902 & 0.082\\\hline\hline

Recurrent neural network & \(y =\) Negative & \(y =\)Positive\\\hline
\(\widetilde{y} =\)Positive & 0.085 & 0.939\\
\(\widetilde{y} =\)Negative & 0.914 & 0.061\\\hline\hline

Non-local aggregation & \(y =\) Negative & \(y =\)Positive\\\hline
\(\widetilde{y} =\)Positive & 0.080 & 0.939\\
\(\widetilde{y} =\)Negative & 0.920 & 0.061\\\hline\hline

Temporal convolution & \(y =\) Negative & \(y =\)Positive\\\hline
\(\widetilde{y} =\)Positive & 0.073 & 0.959\\
\(\widetilde{y} =\)Negative & 0.927 & 0.041\\\hline\hline
\end{tabular}}
\end{table}

\subsection{The video-based model analysis}

{The video-based classification accuracy and testing time are compared in Table 6. The frame-independent aggregation is the fastest method, while the temporal convolution achieves the best accuracy and still has fast processing speed (132ms). We note that without the view classification module and use the ground truth view label, the processing can be 60ms. We also give the confusion matrix of four aggregation methods in the Table 7.}

\begin{table}[t]  \caption{Five-view classification accuracy with different aggregation module. The correct classification of a subject indicates all five views are correctly classified.}\label{tab:5}
\resizebox{1\linewidth}{!}{
\centering
\renewcommand\arraystretch{1.2}

\begin{tabular}{l|c|c|c|c}
\hline\hline
Aggregation & Frame-independent & Non-local & RNN & Temp\\\hline
Accuracy & 0.992 & 0.995 & 0.991 & 0.994\\\hline\hline
\end{tabular}}

\end{table}

\begin{figure}[t]
\begin{center}
\includegraphics[width=1\linewidth]{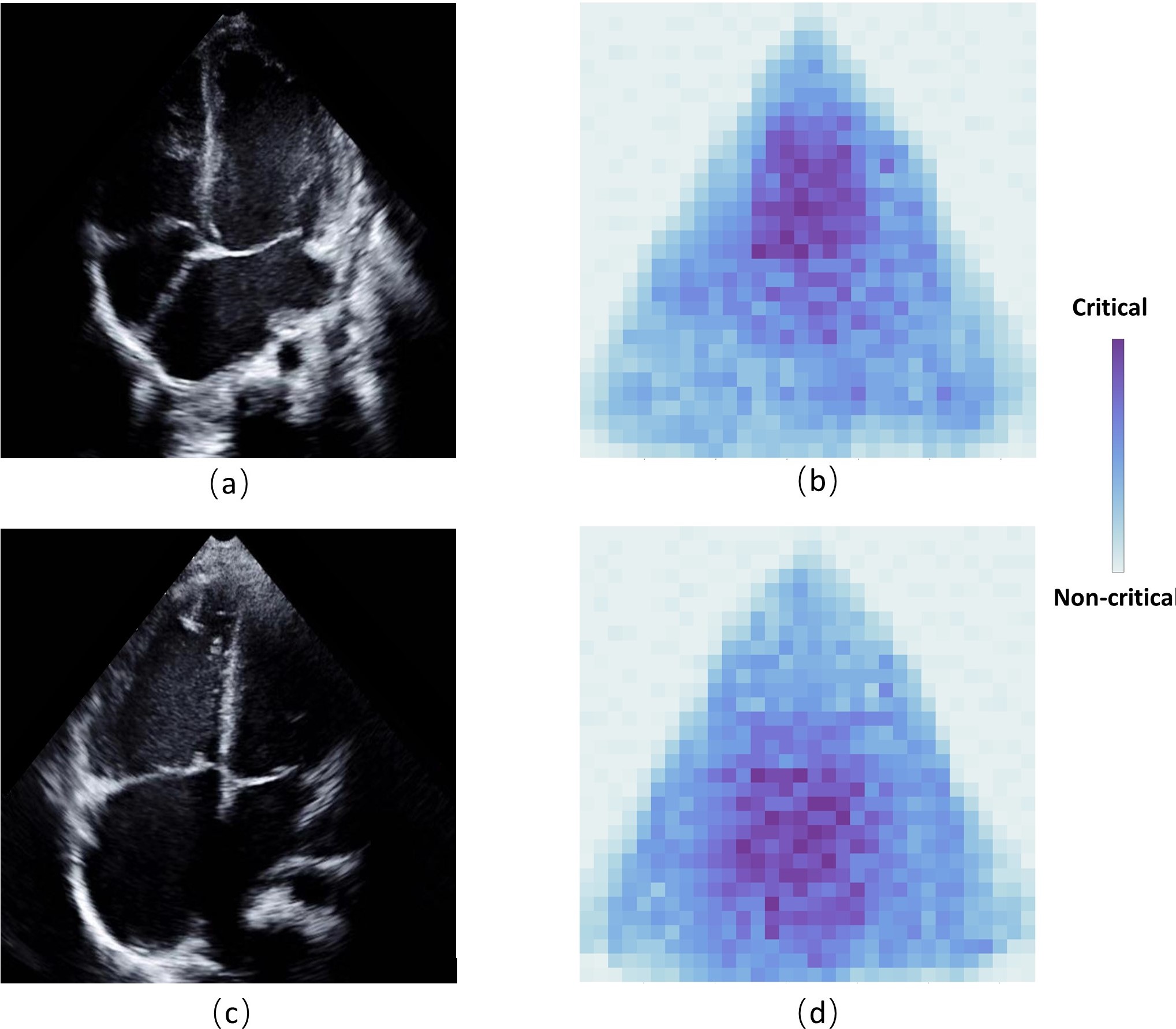}
\end{center} 
\caption{The relative importance of different parts associated with the diagnosis. a. the apical four-chamber view of ASD patient; b. its corresponding relative importance visualization of different parts associated with ASD; c. the apical four-chamber view of VSD patient; d. its corresponding relative importance visualization of different parts associated with VSD.}
\label{fig:12}
\end{figure}

{RNN can outperform the frame-independent aggregation, since it can take the other frames into account. However, its sequential processing can be much slower than the compared methods. The non-local scheme can utilize the powerful parallel computing ability of GPU and achieve comparable or even better performance than RNN with about five times faster processing. The non-local scheme considers all of the positions equally, while the neighboring frames in the echocardiogram video are closely related and follow the cardiac cycle. By considering this property, the temporal convolution can be a good solution and achieves a good balance between the accuracy and processing speed.}

{Moreover, the frame-independent aggregation, RNN, and temporal convolution module can utilize the key-frame label. With the expert labeled keyframe, the performance of the aggregation module can be further improved.} 

{Using the view classification module can achieve comparable performance as using the ground-truth view label. In Table 8, the view detection performance is compared. All of the aggregation methods can well maintain the view information and potentially facilitating the data collection in real-world implementations.}

\begin{figure}[t]
\begin{center}
\includegraphics[width=1\linewidth]{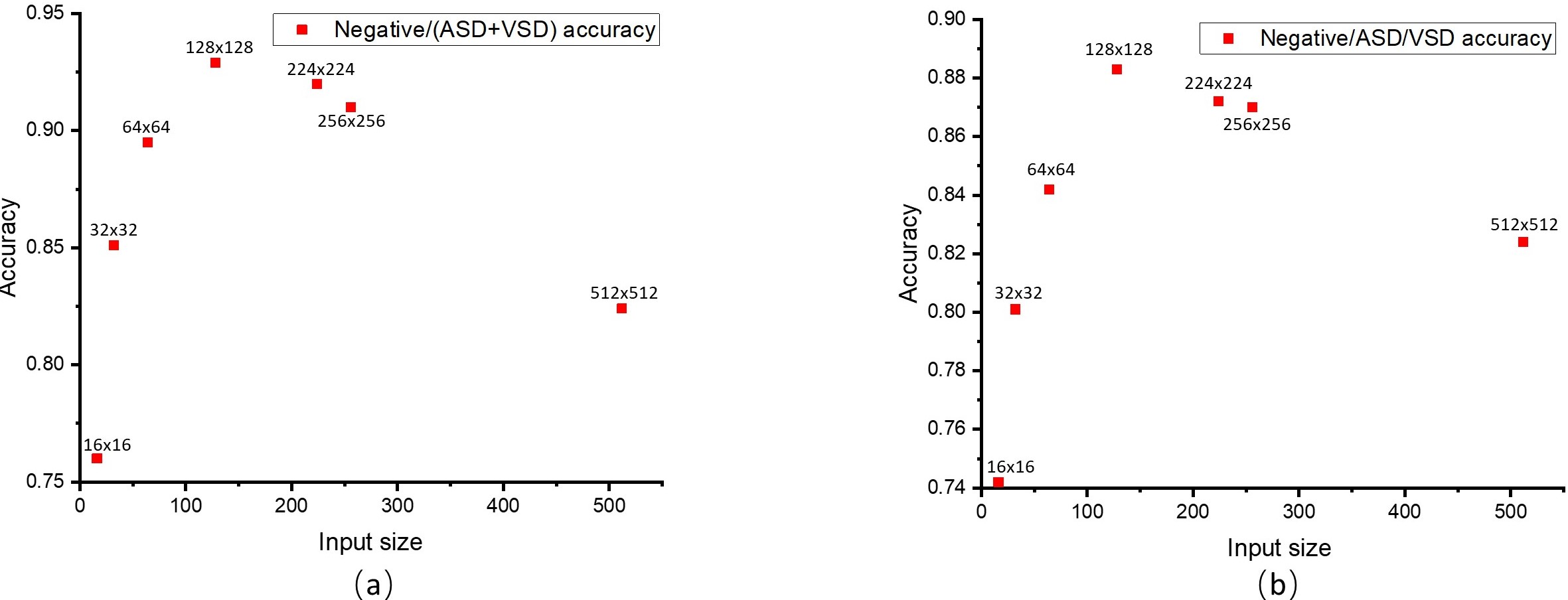}
\end{center} 
\caption{Classification accuracy with different input image sizes. a. binary classification; b. three-class classification.}
\label{fig:13}
\end{figure}

\begin{figure}[t]
\begin{center}
\includegraphics[width=1\linewidth]{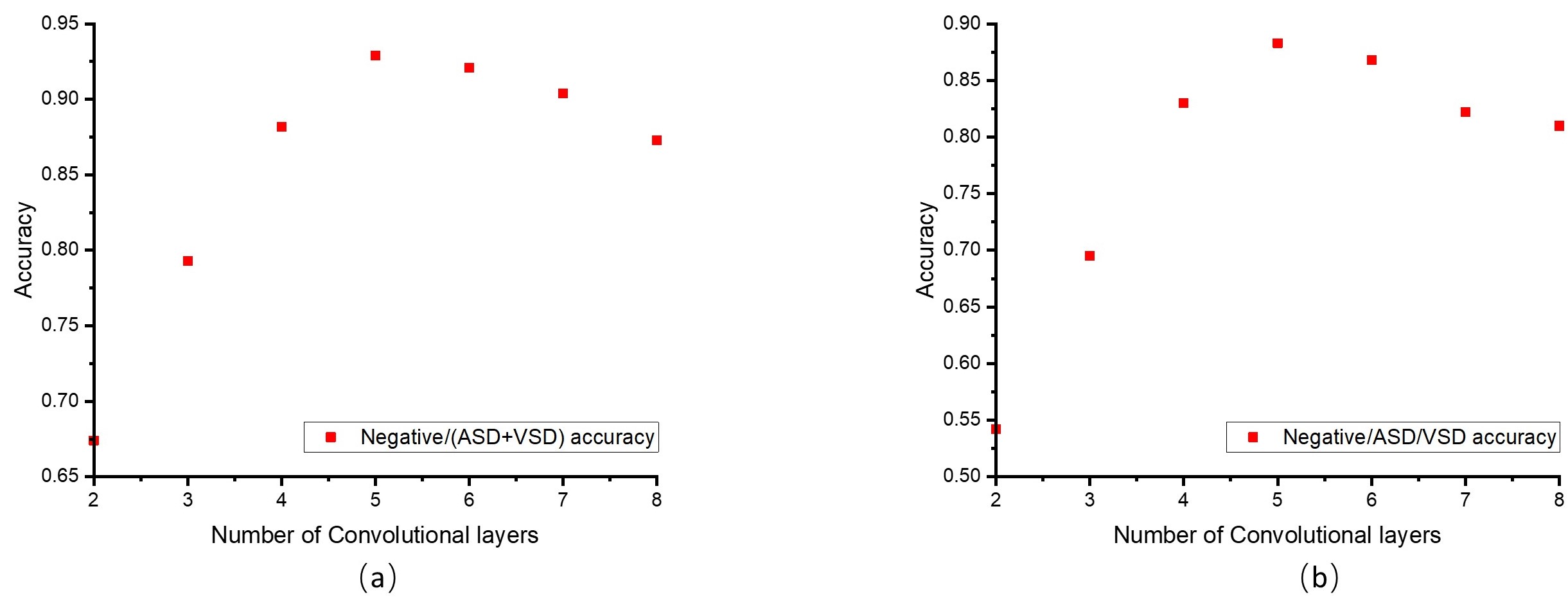}
\end{center} 
\caption{Classification accuracy with different convolutional layers. a. binary classification; b. three-class classification.}
\label{fig:14}
\end{figure}

\subsection{Understanding the perception area}

{Additionally, to understand the behavior of our neural network and identify regions for concatenated input that are critical for medical diagnosis, we propose an image occlusion analysis on our DSC model. We slide a box of 4$\times$4$\times$1 zero-valued pixels along with the layer of the A4C in concatenated ultrasonic images from a patient that was correctly labeled as VSD or ASD samples by our trained model. As a consequence, the importance of each area can be characterized by the relative confidence of the samples being classified as positive. The resulting heat map is shown in Figure \ref{fig:12}. The intensity of the heat map indicates the relative importance of each pixel block. The dark areas decrease the confidence of the model, suggesting that they are critical areas for the diagnosis of cardiopathy. The dark blue regions in Figure \ref{fig:12} coincide with the ROIs of the VSD and ASD in the ultrasonic images, respectively.}

\subsection{Ablation study}

{The input size can significantly affect system performance. We further analyze the validation accuracy when choosing different input sizes. A low-resolution sample with too-small size (e.g., 16$\times$16) cannot offer sufficient information. Meanwhile, a large size sample may require more neural network parameters and training data. The size 128$\times$128$\times5$ consistently achieves the best performance in our binary and three-class classification settings as shown in Figure \ref{fig:13}.

Using more convolution layers can improve the representation ability of neural networks, but the additional parameters require significantly more training data. As shown in Figure \ref{fig:14}, when we evaluate the performance with two to eight convolutional layers, respectively, the five-layer setting can well balance the number of parameters and training samples in our dataset.}

\section{Discussion}

{Clinical diagnosis of CHD is usually based on the comprehensive analysis of multiple views, and the disease diagnosis is a challenging task for the primary doctors. This study intends to develop a practical model to efficiently fuse the information in different views with either image or video modality. Based on clinical experience, we collect the five-view echocardiogram videos from 1308 children, with fine-grained key-frame labels.}

{As we have seen in the results, the 2D key-frame model can diagnose CHD or negative samples with an accuracy of \ 0.954, and in negative, VSD or ASD classification with an accuracy of \ 0.923. Moreover, the video-based model can diagnose with an accuracy of 0.939 (binary classification), and 0.921 (3-class classification) in a collected 2D video testing set, which does not need key-frame selection and view annotation in testing. The present model developed has high diagnostic rates for VSD and ASD that can be potentially applied to the clinical practice in the future.}

\subsection{Clinical insights of multi-view analysis for CHD}

{The multi-view diagnosis accuracy can be 0.05 higher than the diagnose with only the A4C view as shown in Table 4. The AUC metric also coincides with the accuracy. To the best of our knowledge, this is the first effort to explore the five-view echocardiogram, and the corresponding dataset.} 

{Previous researches on the use of artificial intelligence-assisted image processing mainly involved single view research, which can show the application value of artificial intelligence to a certain extent \citep{zheng2008four}. However, the clinical diagnosis of CHD cannot be reliably obtained from a single view. Instead, a multi-view joint diagnosis is necessary. A complete examination requires that the cardiovascular structures be imaged from multiple orthogonal planes. Therefore, we chose five standard 2D views, PSLAX of left ventricle, PSSAX of aorta, A4C, SXLAX of two atria and SSLAX of aortic arch, which in combination could describe all of the major cardiovascular structures in sequence \citep{lai2006guidelines}. This study is the first to use five 2D views to assist with the diagnosis of CHD. Fortunately, our study showed that artificial intelligence and deep learning of the five standard 2D views could make identification and classification diagnosis of CHD.}

{The ultrasound manifestations of CHD can be divided into direct signs and indirect signs. Taking ASD as an example, the direct signs are that the echo is interrupted in the A4C and the SXLAX of two atria, and the blood flow communicates between the two atria by the defect. The indirect signs are that the right ventricle is enlarged in the PSLAX of left ventricle, the right atrium and ventricle are enlarged in the A4C, and the main pulmonary artery is widened in the PSSAX of aorta. For VSD, the direct signs are that the echo is interrupted in the A4C and the PSLAX of left ventricle, and the blood flow communicates between the two ventricles through the defect. The typical VSD indirect signs are that left atrium and ventricle are enlarged in the PSLAX of left ventricle and the A4C. However, when VSD is combined with pulmonary hypertension, it can also manifest as right atrium and ventricle enlargement and pulmonary artery widening which can be easily confused with ASD. At this time, PSLAX of left ventricle and the A4C show the ventricular defect, and the SXLAX of two {atria} show the completed {atria}l septum. These performances can help us to differentiate between the diagnosis of VSD and ASD, thereby showing the importance of multi-view joint diagnosis.}

{Furthermore, the appearance of CHD is various. For example, the defect site of VSD is variable, and defects at different positions will be displayed on different views. The defect can be shown in the A4C when it is located in the muscle and perimembrane, shown in the PSLAX of left ventricle when it is located in the infracristal, and shown in the PSSAX of aorta when it is located in the subarterial, respectively. So comprehensive judgment of multi-view images is required to achieve the correct diagnosis.}

\subsection{Key-frame based multi-view analysis}

{In this work, we have investigated both the multi-channel and multi-branch CNN models with AlexNet and DSC backbones to aggregate the information in different views.}

{The single branch multi-channel structure can efficiently incorporate all the five views and correlate each channel via the pointwise convolution. Compared with the multi-branch model which fuse the information with the simple concatenation, the multi-channel model is possible to adaptively fuse the information in all of the layers.}  

{Targeting for the limited training data, the DSC is adopted which compresses the parameter used in our AlexNet baseline by more than 32 times. To alleviate the class imbalance problem in the diagnosis task, the virtual patients are constructed by randomly sampling the missed view from the different patients as data augmentation. Its lite structure makes it promise to be deployed on many micro-embedded systems at a low cost. With the DSC, data augmentation and transfer learning for either binary classification or three-class classification, our method requires less training data and found excellent performance.}  

{As investigated in our ablation study (Fig. 14), the spatial size of $128\times128$ can be an ideal choice of the input size. The cardiopathy region is usually a hole that requires sufficiently spatial resolution for diagnosing. While the larger input is found to be redundant. The choice of five convolution layer is based on our dataset. With more training data, we may expect a more representative model using deeper networks.}

\subsection{Video-based multi-view analysis}

{The collected raw ultrasonic data is video, i.e., a sequence of images. The key-frame selection is required for the 2D standard view-based solution, which can be tedious in implementation and hard for primary doctors. Actually, the video should cover the information of the key-frame, since the key-frame is a sub-set of the video. To alleviate the manually labeling task, we proposed a soft-attention framework to aggregate the information from these sequential frames.}  

{We investigated four possible aggregation methods to adaptively assign the attention in each frame from the diagnosis perspective rather than the general image quality (e.g., blurry level). We systematically analyzed their performance and processing speed in Table 6.}

{The frame-independent aggregation is the fastest method, while the temporal convolution achieves the best accuracy and still has fast processing speed (132ms). RNN can take the other frames into account, but its sequential processing can be much slower than the other solutions. With parallel computing the non-local scheme can speed up the processing, but it redundantly considers all of the positions equally. However, the echocardiogram video follows the cardiac cycle, in which the frame is closely related to its neighboring frames. The temporal convolution explores the neighboring frames with its proper reception fields and the parallel convolution operation, which achieves the real-time processing without sacrificing the accuracy.}

{With the key-frame annotation in the training samples, the aggregation framework, especially the temporal convolution scheme, can achieve comparable accuracy than the 2D standard view-based solution.} 

{{The five views should follow the fixed order of views} to constitute an input of our model. As shown in Table 8, our multi-task model can detect the view with high accuracy and order them for the diagnosis. Actually, each view has significant patterns that can usually easy to detect. The performance gap of using view classifier and ground-truth view label is marginal in Table 7. The doctors in primary hospitals can collect the videos in random order without the view annotation. Moreover, the view classifier can also be used to double-check the clinical records.}

\subsection{Clinical prospects}

{The model developed in this study has high diagnostic rates for VSD and ASD and can be potentially applied to the clinical practice in the future, especially in primary hospitals. For example, 2D standard views are selected by a doctor, and a preliminary diagnosis of the echocardiogram image can be performed by an end-to-end multi-view deep learning system. Then the positive cases can be referred to specialist hospitals for further treatment. The short-term automated machine learning process can partially replace and promote the long-term professional training of primary doctors, improving the primary diagnosis rate of CHD in China, and laying the foundation for early diagnosis and timely treatment of children with CHD.}

\subsection{Limitations and future directions}

{Our model requires the input of five videos from five pre-defined views. We plan to construct a free-hand diagnosis system by detecting the view of each frame and aggregate the frames from the same view automatically. Moreover, a feature work can be using active learning to instruct the scanning which may reduce the number of views to be collected in real-world implementations.
Thirdly, we will gradually include more echocardiograms of patients with other CHD subtypes in the following study, so that this diagnostic system can complete the diagnosis of more CHD subtypes in the future.}

\section{Conclusion}

{This study proposes to automatically analyze the five-view echocardiograms with end-to-end neural networks. Based on the collected multi-view dataset with both disease labels and standard-view key frame labels, our model can make the diagnosis on either selected 2D standard views or original videos. The model has high diagnostic rates for VSD and ASD and can be potentially applied to the clinical practice in the future, especially in primary hospitals. For example, 2D standard views are selected by a doctor, and a preliminary diagnosis of the echocardiogram image can be performed by artificial intelligence. Then the positive cases can be referred to specialist hospitals for further treatment. The short-term automated machine learning process can partially replace and promote the long-term professional training of primary doctors, improving the primary diagnosis rate of CHD in China, and laying the foundation for early diagnosis and timely treatment of children with CHD.}

\section*{Acknowledgements}

This work was supported by the National Natural Science Foundation of China [grant number 91846102], the Fundamental Research Funds for the Central Universities [grant number GK2240260006], Beijing Municipal Administration of Hospitals Youth Programme [grant number QML20191208] and Jiangsu Youth Programme [grant number BK20200238].
%%Harvard
\bibliographystyle{model2-names.bst}\biboptions{authoryear}
\bibliography{refs}

\begin{thebibliography}{59}
\expandafter\ifx\csname natexlab\endcsname\relax\def\natexlab#1{#1}\fi
\providecommand{\url}[1]{\texttt{#1}}
\providecommand{\href}[2]{#2}
\providecommand{\path}[1]{#1}
\providecommand{\DOIprefix}{doi:}
\providecommand{\ArXivprefix}{arXiv:}
\providecommand{\URLprefix}{URL: }
\providecommand{\Pubmedprefix}{pmid:}
\providecommand{\doi}[1]{\href{http://dx.doi.org/#1}{\path{#1}}}
\providecommand{\Pubmed}[1]{\href{pmid:#1}{\path{#1}}}
\providecommand{\bibinfo}[2]{#2}
\ifx\xfnm\relax \def\xfnm[#1]{\unskip,\space#1}\fi
%Type = Article
\bibitem[{Bai et~al.(2018)Bai, Kolter and Koltun}]{bai2018empirical}
\bibinfo{author}{Bai, S.}, \bibinfo{author}{Kolter, J.Z.}, \bibinfo{author}{Koltun, V.}, \bibinfo{year}{2018}.
\newblock \bibinfo{title}{An empirical evaluation of generic convolutional and recurrent networks for sequence modeling}.
\newblock \bibinfo{journal}{arXiv preprint arXiv:1803.01271} .
%Type = Inproceedings
\bibitem[{Battaglia et~al.(2016)Battaglia, Pascanu, Lai, Rezende et~al.}]{battaglia2016interaction}
\bibinfo{author}{Battaglia, P.}, \bibinfo{author}{Pascanu, R.}, \bibinfo{author}{Lai, M.}, \bibinfo{author}{Rezende, D.J.}, et~al., \bibinfo{year}{2016}.
\newblock \bibinfo{title}{Interaction networks for learning about objects, relations and physics}, in: \bibinfo{booktitle}{Advances in neural information processing systems}, pp. \bibinfo{pages}{4502--4510}.
%Type = Article
\bibitem[{Bejnordi et~al.(2017)Bejnordi, Veta, Van~Diest, Van~Ginneken, Karssemeijer, Litjens, Van Der~Laak, Hermsen, Manson, Balkenhol et~al.}]{bejnordi2017diagnostic}
\bibinfo{author}{Bejnordi, B.E.}, \bibinfo{author}{Veta, M.}, \bibinfo{author}{Van~Diest, P.J.}, \bibinfo{author}{Van~Ginneken, B.}, \bibinfo{author}{Karssemeijer, N.}, \bibinfo{author}{Litjens, G.}, \bibinfo{author}{Van Der~Laak, J.A.}, \bibinfo{author}{Hermsen, M.}, \bibinfo{author}{Manson, Q.F.}, \bibinfo{author}{Balkenhol, M.}, et~al., \bibinfo{year}{2017}.
\newblock \bibinfo{title}{Diagnostic assessment of deep learning algorithms for detection of lymph node metastases in women with breast cancer}.
\newblock \bibinfo{journal}{Jama} \bibinfo{volume}{318}, \bibinfo{pages}{2199--2210}.
%Type = Inproceedings
\bibitem[{Buades et~al.(2005)Buades, Coll and Morel}]{buades2005non}
\bibinfo{author}{Buades, A.}, \bibinfo{author}{Coll, B.}, \bibinfo{author}{Morel, J.M.}, \bibinfo{year}{2005}.
\newblock \bibinfo{title}{A non-local algorithm for image denoising}, in: \bibinfo{booktitle}{Computer Vision and Pattern Recognition, 2005. CVPR 2005. IEEE Computer Society Conference on}, \bibinfo{organization}{IEEE}. pp. \bibinfo{pages}{60--65}.
%Type = Article
\bibitem[{Chang et~al.(2008)Chang, Gurvitz and Rodriguez}]{chang2008missed}
\bibinfo{author}{Chang, R.K.R.}, \bibinfo{author}{Gurvitz, M.}, \bibinfo{author}{Rodriguez, S.}, \bibinfo{year}{2008}.
\newblock \bibinfo{title}{Missed diagnosis of critical congenital heart disease}.
\newblock \bibinfo{journal}{Archives of pediatrics \& adolescent medicine} \bibinfo{volume}{162}, \bibinfo{pages}{969--974}.
%Type = Article
\bibitem[{Che et~al.(2019)Che, Liu, Li, Ge, Zhang, Xiong and Bengio}]{che2019deep}
\bibinfo{author}{Che, T.}, \bibinfo{author}{Liu, X.}, \bibinfo{author}{Li, S.}, \bibinfo{author}{Ge, Y.}, \bibinfo{author}{Zhang, R.}, \bibinfo{author}{Xiong, C.}, \bibinfo{author}{Bengio, Y.}, \bibinfo{year}{2019}.
\newblock \bibinfo{title}{Deep verifier networks: Verification of deep discriminative models with deep generative models}.
\newblock \bibinfo{journal}{arXiv preprint arXiv:1911.07421} .
%Type = Article
\bibitem[{Cortes and Vapnik(1995)}]{cortes1995support}
\bibinfo{author}{Cortes, C.}, \bibinfo{author}{Vapnik, V.}, \bibinfo{year}{1995}.
\newblock \bibinfo{title}{Support-vector networks}.
\newblock \bibinfo{journal}{Machine learning} \bibinfo{volume}{20}, \bibinfo{pages}{273--297}.
%Type = Inproceedings
\bibitem[{Criminisi et~al.(2009)Criminisi, Shotton and Bucciarelli}]{criminisi2009decision}
\bibinfo{author}{Criminisi, A.}, \bibinfo{author}{Shotton, J.}, \bibinfo{author}{Bucciarelli, S.}, \bibinfo{year}{2009}.
\newblock \bibinfo{title}{Decision forests with long-range spatial context for organ localization in ct volumes}, in: \bibinfo{booktitle}{Medical Image Computing and Computer-Assisted Intervention (MICCAI)}, pp. \bibinfo{pages}{69--80}.
%Type = Article
\bibitem[{Dai et~al.(2011)Dai, Zhu, Liang, Wang, Wang and Mao}]{dai2011birth}
\bibinfo{author}{Dai, L.}, \bibinfo{author}{Zhu, J.}, \bibinfo{author}{Liang, J.}, \bibinfo{author}{Wang, Y.P.}, \bibinfo{author}{Wang, H.}, \bibinfo{author}{Mao, M.}, \bibinfo{year}{2011}.
\newblock \bibinfo{title}{Birth defects surveillance in china}.
\newblock \bibinfo{journal}{World journal of pediatrics} \bibinfo{volume}{7}, \bibinfo{pages}{302}.
%Type = Article
\bibitem[{De~Fauw et~al.(2018)De~Fauw, Ledsam, Romera-Paredes, Nikolov, Tomasev, Blackwell, Askham, Glorot, O’Donoghue, Visentin et~al.}]{de2018clinically}
\bibinfo{author}{De~Fauw, J.}, \bibinfo{author}{Ledsam, J.R.}, \bibinfo{author}{Romera-Paredes, B.}, \bibinfo{author}{Nikolov, S.}, \bibinfo{author}{Tomasev, N.}, \bibinfo{author}{Blackwell, S.}, \bibinfo{author}{Askham, H.}, \bibinfo{author}{Glorot, X.}, \bibinfo{author}{O’Donoghue, B.}, \bibinfo{author}{Visentin, D.}, et~al., \bibinfo{year}{2018}.
\newblock \bibinfo{title}{Clinically applicable deep learning for diagnosis and referral in retinal disease}.
\newblock \bibinfo{journal}{Nature medicine} \bibinfo{volume}{24}, \bibinfo{pages}{1342--1350}.
%Type = Article
\bibitem[{Diller et~al.(2019)Diller, Babu-Narayan, Li, Radojevic, Kempny, Uebing, Dimopoulos, Baumgartner, Gatzoulis and Orwat}]{diller2019utility}
\bibinfo{author}{Diller, G.P.}, \bibinfo{author}{Babu-Narayan, S.}, \bibinfo{author}{Li, W.}, \bibinfo{author}{Radojevic, J.}, \bibinfo{author}{Kempny, A.}, \bibinfo{author}{Uebing, A.}, \bibinfo{author}{Dimopoulos, K.}, \bibinfo{author}{Baumgartner, H.}, \bibinfo{author}{Gatzoulis, M.A.}, \bibinfo{author}{Orwat, S.}, \bibinfo{year}{2019}.
\newblock \bibinfo{title}{Utility of machine learning algorithms in assessing patients with a systemic right ventricle}.
\newblock \bibinfo{journal}{European Heart Journal-Cardiovascular Imaging} \bibinfo{volume}{20}, \bibinfo{pages}{925--931}.
%Type = Article
\bibitem[{Esteva et~al.(2017)Esteva, Kuprel, Novoa, Ko, Swetter, Blau and Thrun}]{esteva2017dermatologist}
\bibinfo{author}{Esteva, A.}, \bibinfo{author}{Kuprel, B.}, \bibinfo{author}{Novoa, R.A.}, \bibinfo{author}{Ko, J.}, \bibinfo{author}{Swetter, S.M.}, \bibinfo{author}{Blau, H.M.}, \bibinfo{author}{Thrun, S.}, \bibinfo{year}{2017}.
\newblock \bibinfo{title}{Dermatologist-level classification of skin cancer with deep neural networks}.
\newblock \bibinfo{journal}{Nature} \bibinfo{volume}{542}, \bibinfo{pages}{115--118}.
%Type = Article
\bibitem[{Gao and Nevatia(2018)}]{gao2018revisiting}
\bibinfo{author}{Gao, J.}, \bibinfo{author}{Nevatia, R.}, \bibinfo{year}{2018}.
\newblock \bibinfo{title}{Revisiting temporal modeling for video-based person reid}.
\newblock \bibinfo{journal}{arXiv preprint arXiv:1805.02104} .
%Type = Inproceedings
\bibitem[{Han et~al.(2020)Han, Liu, Sheng, Ren, Han, You, Liu and Luo}]{Han_2020_CVPR_Workshops}
\bibinfo{author}{Han, Y.}, \bibinfo{author}{Liu, X.}, \bibinfo{author}{Sheng, Z.}, \bibinfo{author}{Ren, Y.}, \bibinfo{author}{Han, X.}, \bibinfo{author}{You, J.}, \bibinfo{author}{Liu, R.}, \bibinfo{author}{Luo, Z.}, \bibinfo{year}{2020}.
\newblock \bibinfo{title}{Wasserstein loss-based deep object detection}, in: \bibinfo{booktitle}{Proceedings of the IEEE/CVF Conference on Computer Vision and Pattern Recognition (CVPR) Workshops}.
%Type = Inproceedings
\bibitem[{He et~al.(2020)He, Liu, Fan and You}]{He_2020_CVPR_Workshops}
\bibinfo{author}{He, G.}, \bibinfo{author}{Liu, X.}, \bibinfo{author}{Fan, F.}, \bibinfo{author}{You, J.}, \bibinfo{year}{2020}.
\newblock \bibinfo{title}{Image2audio: Facilitating semi-supervised audio emotion recognition with facial expression image}, in: \bibinfo{booktitle}{Proceedings of the IEEE/CVF Conference on Computer Vision and Pattern Recognition (CVPR) Workshops}.
%Type = Article
\bibitem[{Henderson et~al.(2017)Henderson, Islam, Bachman, Pineau, Precup and Meger}]{henderson2017deep}
\bibinfo{author}{Henderson, P.}, \bibinfo{author}{Islam, R.}, \bibinfo{author}{Bachman, P.}, \bibinfo{author}{Pineau, J.}, \bibinfo{author}{Precup, D.}, \bibinfo{author}{Meger, D.}, \bibinfo{year}{2017}.
\newblock \bibinfo{title}{Deep reinforcement learning that matters}.
\newblock \bibinfo{journal}{arXiv preprint arXiv:1709.06560} .
%Type = Inproceedings
\bibitem[{Hoshen(2017)}]{hoshen2017vain}
\bibinfo{author}{Hoshen, Y.}, \bibinfo{year}{2017}.
\newblock \bibinfo{title}{Vain: Attentional multi-agent predictive modeling}, in: \bibinfo{booktitle}{Advances in Neural Information Processing Systems}, pp. \bibinfo{pages}{2701--2711}.
%Type = Article
\bibitem[{Howard et~al.(2017)Howard, Zhu, Chen, Kalenichenko, Wang, Weyand, Andreetto and Adam}]{howard2017mobilenets}
\bibinfo{author}{Howard, A.G.}, \bibinfo{author}{Zhu, M.}, \bibinfo{author}{Chen, B.}, \bibinfo{author}{Kalenichenko, D.}, \bibinfo{author}{Wang, W.}, \bibinfo{author}{Weyand, T.}, \bibinfo{author}{Andreetto, M.}, \bibinfo{author}{Adam, H.}, \bibinfo{year}{2017}.
\newblock \bibinfo{title}{Mobilenets: Efficient convolutional neural networks for mobile vision applications}.
\newblock \bibinfo{journal}{arXiv preprint arXiv:1704.04861} .
%Type = Article
\bibitem[{Kwitt et~al.(2013)Kwitt, Vasconcelos, Razzaque and Aylward}]{kwitt2013localizing}
\bibinfo{author}{Kwitt, R.}, \bibinfo{author}{Vasconcelos, N.}, \bibinfo{author}{Razzaque, S.}, \bibinfo{author}{Aylward, S.}, \bibinfo{year}{2013}.
\newblock \bibinfo{title}{Localizing target structures in ultrasound video--a phantom study}.
\newblock \bibinfo{journal}{Medical image analysis} \bibinfo{volume}{17}, \bibinfo{pages}{712--722}.
%Type = Article
\bibitem[{Lai et~al.(2006)Lai, Geva, Shirali, Frommelt, Humes, Brook, Pignatelli and Rychik}]{lai2006guidelines}
\bibinfo{author}{Lai, W.W.}, \bibinfo{author}{Geva, T.}, \bibinfo{author}{Shirali, G.S.}, \bibinfo{author}{Frommelt, P.C.}, \bibinfo{author}{Humes, R.A.}, \bibinfo{author}{Brook, M.M.}, \bibinfo{author}{Pignatelli, R.H.}, \bibinfo{author}{Rychik, J.}, \bibinfo{year}{2006}.
\newblock \bibinfo{title}{Guidelines and standards for performance of a pediatric echocardiogram: a report from the task force of the pediatric council of the american society of echocardiography}.
\newblock \bibinfo{journal}{Journal of the American Society of Echocardiography} \bibinfo{volume}{19}, \bibinfo{pages}{1413--1430}.
%Type = Article
\bibitem[{LeCun et~al.(2015)LeCun, Bengio and Hinton}]{lecun2015deep}
\bibinfo{author}{LeCun, Y.}, \bibinfo{author}{Bengio, Y.}, \bibinfo{author}{Hinton, G.}, \bibinfo{year}{2015}.
\newblock \bibinfo{title}{Deep learning}.
\newblock \bibinfo{journal}{nature} \bibinfo{volume}{521}, \bibinfo{pages}{436--444}.
%Type = Inproceedings
\bibitem[{Lee et~al.(2016)Lee, Lee and Kim}]{lee2016multi}
\bibinfo{author}{Lee, D.}, \bibinfo{author}{Lee, J.}, \bibinfo{author}{Kim, K.E.}, \bibinfo{year}{2016}.
\newblock \bibinfo{title}{Multi-view automatic lip-reading using neural network}, in: \bibinfo{booktitle}{Asian conference on computer vision}, \bibinfo{organization}{Springer}. pp. \bibinfo{pages}{290--302}.
%Type = Article
\bibitem[{van~der Linde et~al.(2011)van~der Linde, Konings, Slager, Witsenburg, Helbing, Takkenberg and Roos-Hesselink}]{van2011birth}
\bibinfo{author}{van~der Linde, D.}, \bibinfo{author}{Konings, E.E.}, \bibinfo{author}{Slager, M.A.}, \bibinfo{author}{Witsenburg, M.}, \bibinfo{author}{Helbing, W.A.}, \bibinfo{author}{Takkenberg, J.J.}, \bibinfo{author}{Roos-Hesselink, J.W.}, \bibinfo{year}{2011}.
\newblock \bibinfo{title}{Birth prevalence of congenital heart disease worldwide: a systematic review and meta-analysis}.
\newblock \bibinfo{journal}{Journal of the American College of Cardiology} \bibinfo{volume}{58}, \bibinfo{pages}{2241--2247}.
%Type = Article
\bibitem[{Litjens et~al.(2019)Litjens, Ciompi, Wolterink, de~Vos, Leiner, Teuwen and I{\v{s}}gum}]{litjens2019state}
\bibinfo{author}{Litjens, G.}, \bibinfo{author}{Ciompi, F.}, \bibinfo{author}{Wolterink, J.M.}, \bibinfo{author}{de~Vos, B.D.}, \bibinfo{author}{Leiner, T.}, \bibinfo{author}{Teuwen, J.}, \bibinfo{author}{I{\v{s}}gum, I.}, \bibinfo{year}{2019}.
\newblock \bibinfo{title}{State-of-the-art deep learning in cardiovascular image analysis}.
\newblock \bibinfo{journal}{JACC: Cardiovascular Imaging} \bibinfo{volume}{12}, \bibinfo{pages}{1549--1565}.
%Type = Article
\bibitem[{Liu et~al.(2019a)Liu, Wang, Yang, Lei, Liu, Li, Ni and Wang}]{liu2019deep}
\bibinfo{author}{Liu, S.}, \bibinfo{author}{Wang, Y.}, \bibinfo{author}{Yang, X.}, \bibinfo{author}{Lei, B.}, \bibinfo{author}{Liu, L.}, \bibinfo{author}{Li, S.X.}, \bibinfo{author}{Ni, D.}, \bibinfo{author}{Wang, T.}, \bibinfo{year}{2019}a.
\newblock \bibinfo{title}{Deep learning in medical ultrasound analysis: a review}.
\newblock \bibinfo{journal}{Engineering} .
%Type = Article
\bibitem[{Liu(2020)}]{liu2020disentanglement}
\bibinfo{author}{Liu, X.}, \bibinfo{year}{2020}.
\newblock \bibinfo{title}{Disentanglement for discriminative visual recognition}.
\newblock \bibinfo{journal}{arXiv preprint arXiv:2006.07810} .
%Type = Inproceedings
\bibitem[{Liu et~al.(2018a)Liu, B.V.K, Yang, Tang and You}]{liu2018dependency}
\bibinfo{author}{Liu, X.}, \bibinfo{author}{B.V.K, K.}, \bibinfo{author}{Yang, C.}, \bibinfo{author}{Tang, Q.}, \bibinfo{author}{You, J.}, \bibinfo{year}{2018}a.
\newblock \bibinfo{title}{Dependency-aware attention control for unconstrained face recognition with image sets}, in: \bibinfo{booktitle}{European Conference on Computer Vision}.
%Type = Article
\bibitem[{Liu et~al.(2020a)Liu, Fan, Kong, Xie, Lu and You}]{liu2020unimodal}
\bibinfo{author}{Liu, X.}, \bibinfo{author}{Fan, F.}, \bibinfo{author}{Kong, L.}, \bibinfo{author}{Xie, W.}, \bibinfo{author}{Lu, J.}, \bibinfo{author}{You, J.}, \bibinfo{year}{2020}a.
\newblock \bibinfo{title}{Unimodal regularized neuron stick-breaking for ordinal classification}.
\newblock \bibinfo{journal}{Neurocomputing} .
%Type = Inproceedings
\bibitem[{Liu et~al.(2019b)Liu, Guo, Jia and Kumar}]{liu2019dependency}
\bibinfo{author}{Liu, X.}, \bibinfo{author}{Guo, Z.}, \bibinfo{author}{Jia, J.}, \bibinfo{author}{Kumar, B.}, \bibinfo{year}{2019}b.
\newblock \bibinfo{title}{Dependency-aware attention control for imageset-based face recognition}, in: \bibinfo{booktitle}{ArXiv}.
%Type = Inproceedings
\bibitem[{Liu et~al.(2019c)Liu, Guo, Li, You and B.V.K}]{liu2019permutation}
\bibinfo{author}{Liu, X.}, \bibinfo{author}{Guo, Z.}, \bibinfo{author}{Li, S.}, \bibinfo{author}{You, J.}, \bibinfo{author}{B.V.K, K.}, \bibinfo{year}{2019}c.
\newblock \bibinfo{title}{Dependency-aware attention control for unconstrained face recognition with image sets}, in: \bibinfo{booktitle}{ICCV}.
%Type = Inproceedings
\bibitem[{Liu et~al.(2019d)Liu, Han, Qiao, Ge, Li and Lu}]{liu2019unimodal}
\bibinfo{author}{Liu, X.}, \bibinfo{author}{Han, X.}, \bibinfo{author}{Qiao, Y.}, \bibinfo{author}{Ge, Y.}, \bibinfo{author}{Li, S.}, \bibinfo{author}{Lu, J.}, \bibinfo{year}{2019}d.
\newblock \bibinfo{title}{Unimodal-uniform constrained wasserstein training for medical diagnosis}, in: \bibinfo{booktitle}{Proceedings of the IEEE International Conference on Computer Vision Workshops}, pp. \bibinfo{pages}{0--0}.
%Type = Inproceedings
\bibitem[{Liu et~al.(2020b)Liu, Xing, Yang, Kuo, El~Fakhri and Woo}]{liu2020symmetric}
\bibinfo{author}{Liu, X.}, \bibinfo{author}{Xing, F.}, \bibinfo{author}{Yang, C.}, \bibinfo{author}{Kuo, C.J.}, \bibinfo{author}{El~Fakhri, G.}, \bibinfo{author}{Woo, J.}, \bibinfo{year}{2020}b.
\newblock \bibinfo{title}{Symmetric-constrained irregular structure inpainting for brain mri registration with tumor pathology}, in: \bibinfo{booktitle}{MICCAI BrainLes}.
%Type = Inproceedings
\bibitem[{Liu et~al.(2019e)Liu, Zou, Che, Ding, Jia, You and Kumar}]{Liu_2019_ICCV}
\bibinfo{author}{Liu, X.}, \bibinfo{author}{Zou, Y.}, \bibinfo{author}{Che, T.}, \bibinfo{author}{Ding, P.}, \bibinfo{author}{Jia, P.}, \bibinfo{author}{You, J.}, \bibinfo{author}{Kumar, B.V.}, \bibinfo{year}{2019}e.
\newblock \bibinfo{title}{Conservative wasserstein training for pose estimation}, in: \bibinfo{booktitle}{Proceedings of the IEEE/CVF International Conference on Computer Vision (ICCV)}.
%Type = Inproceedings
\bibitem[{Liu et~al.(2018b)Liu, Zou, Song, Yang, You and K~Vijaya~Kumar}]{liu2018ordinal}
\bibinfo{author}{Liu, X.}, \bibinfo{author}{Zou, Y.}, \bibinfo{author}{Song, Y.}, \bibinfo{author}{Yang, C.}, \bibinfo{author}{You, J.}, \bibinfo{author}{K~Vijaya~Kumar, B.}, \bibinfo{year}{2018}b.
\newblock \bibinfo{title}{Ordinal regression with neuron stick-breaking for medical diagnosis}, in: \bibinfo{booktitle}{Proceedings of the European Conference on Computer Vision (ECCV)}, pp. \bibinfo{pages}{0--0}.
%Type = Inproceedings
\bibitem[{Liu et~al.(2017)Liu, Yan and Ouyang}]{liu2017quality}
\bibinfo{author}{Liu, Y.}, \bibinfo{author}{Yan, J.}, \bibinfo{author}{Ouyang, W.}, \bibinfo{year}{2017}.
\newblock \bibinfo{title}{Quality aware network for set to set recognition}, in: \bibinfo{booktitle}{Proc. IEEE Int. Conf. Comput. Vision Pattern Recognit.}, pp. \bibinfo{pages}{5790--5799}.
%Type = Article
\bibitem[{Luo et~al.(2019)Luo, Qin, Wang, Ye, Pan, Huang, Luo, Guo, Peng and Wang}]{luo2019outcomes}
\bibinfo{author}{Luo, H.}, \bibinfo{author}{Qin, G.}, \bibinfo{author}{Wang, L.}, \bibinfo{author}{Ye, Z.}, \bibinfo{author}{Pan, Y.}, \bibinfo{author}{Huang, L.}, \bibinfo{author}{Luo, W.}, \bibinfo{author}{Guo, Q.}, \bibinfo{author}{Peng, Y.}, \bibinfo{author}{Wang, E.}, \bibinfo{year}{2019}.
\newblock \bibinfo{title}{Outcomes of infant cardiac surgery for congenital heart disease concomitant with persistent pneumonia: A retrospective cohort study}.
\newblock \bibinfo{journal}{Journal of cardiothoracic and vascular anesthesia} \bibinfo{volume}{33}, \bibinfo{pages}{428--432}.
%Type = Article
\bibitem[{Maraci et~al.(2017)Maraci, Bridge, Napolitano, Papageorghiou and Noble}]{maraci2017framework}
\bibinfo{author}{Maraci, M.A.}, \bibinfo{author}{Bridge, C.P.}, \bibinfo{author}{Napolitano, R.}, \bibinfo{author}{Papageorghiou, A.}, \bibinfo{author}{Noble, J.A.}, \bibinfo{year}{2017}.
\newblock \bibinfo{title}{A framework for analysis of linear ultrasound videos to detect fetal presentation and heartbeat}.
\newblock \bibinfo{journal}{Medical image analysis} \bibinfo{volume}{37}, \bibinfo{pages}{22--36}.
%Type = Inproceedings
\bibitem[{Maraci et~al.(2018)Maraci, Xie and Noble}]{maraci2018can}
\bibinfo{author}{Maraci, M.A.}, \bibinfo{author}{Xie, W.}, \bibinfo{author}{Noble, J.A.}, \bibinfo{year}{2018}.
\newblock \bibinfo{title}{Can dilated convolutions capture ultrasound video dynamics?}, in: \bibinfo{booktitle}{International Workshop on Machine Learning in Medical Imaging}, \bibinfo{organization}{Springer}. pp. \bibinfo{pages}{116--124}.
%Type = Article
\bibitem[{Pereira et~al.(2017)Pereira, Bueno, Rodriguez, Perrin, Marx, Cardinale, Salgo and del Nido}]{pereira2017automated}
\bibinfo{author}{Pereira, F.}, \bibinfo{author}{Bueno, A.}, \bibinfo{author}{Rodriguez, A.}, \bibinfo{author}{Perrin, D.}, \bibinfo{author}{Marx, G.}, \bibinfo{author}{Cardinale, M.}, \bibinfo{author}{Salgo, I.}, \bibinfo{author}{del Nido, P.}, \bibinfo{year}{2017}.
\newblock \bibinfo{title}{Automated detection of coarctation of aorta in neonates from two-dimensional echocardiograms}.
\newblock \bibinfo{journal}{Journal of Medical Imaging} \bibinfo{volume}{4}, \bibinfo{pages}{014502}.
%Type = Inproceedings
\bibitem[{Pruetz et~al.(2019)Pruetz, Wang and Noori}]{pruetz2019delivery}
\bibinfo{author}{Pruetz, J.D.}, \bibinfo{author}{Wang, S.S.}, \bibinfo{author}{Noori, S.}, \bibinfo{year}{2019}.
\newblock \bibinfo{title}{Delivery room emergencies in critical congenital heart diseases}, in: \bibinfo{booktitle}{Seminars in Fetal and Neonatal Medicine}, \bibinfo{organization}{Elsevier}. p. \bibinfo{pages}{101034}.
%Type = Article
\bibitem[{Simonyan and Zisserman(2014)}]{simonyan2014very}
\bibinfo{author}{Simonyan, K.}, \bibinfo{author}{Zisserman, A.}, \bibinfo{year}{2014}.
\newblock \bibinfo{title}{Very deep convolutional networks for large-scale image recognition}.
\newblock \bibinfo{journal}{arXiv preprint arXiv:1409.1556} .
%Type = Article
\bibitem[{Sun et~al.(2015)Sun, Liu, Lu, Zheng and Zhang}]{sun2015congenital}
\bibinfo{author}{Sun, R.}, \bibinfo{author}{Liu, M.}, \bibinfo{author}{Lu, L.}, \bibinfo{author}{Zheng, Y.}, \bibinfo{author}{Zhang, P.}, \bibinfo{year}{2015}.
\newblock \bibinfo{title}{Congenital heart disease: causes, diagnosis, symptoms, and treatments}.
\newblock \bibinfo{journal}{Cell biochemistry and biophysics} \bibinfo{volume}{72}, \bibinfo{pages}{857--860}.
%Type = Inproceedings
\bibitem[{Szegedy et~al.(2016)Szegedy, Vanhoucke, Ioffe, Shlens and Wojna}]{szegedy2016rethinking}
\bibinfo{author}{Szegedy, C.}, \bibinfo{author}{Vanhoucke, V.}, \bibinfo{author}{Ioffe, S.}, \bibinfo{author}{Shlens, J.}, \bibinfo{author}{Wojna, Z.}, \bibinfo{year}{2016}.
\newblock \bibinfo{title}{Rethinking the inception architecture for computer vision}, in: \bibinfo{booktitle}{Proceedings of the IEEE Conference on Computer Vision and Pattern Recognition}, pp. \bibinfo{pages}{2818--2826}.
%Type = Inproceedings
\bibitem[{Tran et~al.(2015)Tran, Bourdev, Fergus, Torresani and Paluri}]{tran2015learning}
\bibinfo{author}{Tran, D.}, \bibinfo{author}{Bourdev, L.}, \bibinfo{author}{Fergus, R.}, \bibinfo{author}{Torresani, L.}, \bibinfo{author}{Paluri, M.}, \bibinfo{year}{2015}.
\newblock \bibinfo{title}{Learning spatiotemporal features with 3d convolutional networks}, in: \bibinfo{booktitle}{Proceedings of the IEEE international conference on computer vision}, pp. \bibinfo{pages}{4489--4497}.
%Type = Inproceedings
\bibitem[{Vaswani et~al.(2017)Vaswani, Shazeer, Parmar, Uszkoreit, Jones, Gomez, Kaiser and Polosukhin}]{vaswani2017attention}
\bibinfo{author}{Vaswani, A.}, \bibinfo{author}{Shazeer, N.}, \bibinfo{author}{Parmar, N.}, \bibinfo{author}{Uszkoreit, J.}, \bibinfo{author}{Jones, L.}, \bibinfo{author}{Gomez, A.N.}, \bibinfo{author}{Kaiser, {\L}.}, \bibinfo{author}{Polosukhin, I.}, \bibinfo{year}{2017}.
\newblock \bibinfo{title}{Attention is all you need}, in: \bibinfo{booktitle}{Advances in Neural Information Processing Systems}, pp. \bibinfo{pages}{5998--6008}.
%Type = Inproceedings
\bibitem[{Wang et~al.(2018)Wang, Girshick, Gupta and He}]{wang2018non}
\bibinfo{author}{Wang, X.}, \bibinfo{author}{Girshick, R.}, \bibinfo{author}{Gupta, A.}, \bibinfo{author}{He, K.}, \bibinfo{year}{2018}.
\newblock \bibinfo{title}{Non-local neural networks}, in: \bibinfo{booktitle}{The IEEE Conference on Computer Vision and Pattern Recognition (CVPR)}.
%Type = Inproceedings
\bibitem[{Watters et~al.(2017)Watters, Zoran, Weber, Battaglia, Pascanu and Tacchetti}]{watters2017visual}
\bibinfo{author}{Watters, N.}, \bibinfo{author}{Zoran, D.}, \bibinfo{author}{Weber, T.}, \bibinfo{author}{Battaglia, P.}, \bibinfo{author}{Pascanu, R.}, \bibinfo{author}{Tacchetti, A.}, \bibinfo{year}{2017}.
\newblock \bibinfo{title}{Visual interaction networks: Learning a physics simulator from video}, in: \bibinfo{booktitle}{Advances in Neural Information Processing Systems}, pp. \bibinfo{pages}{4539--4547}.
%Type = Article
\bibitem[{WU et~al.(2006)WU, L{\"U} and Liu}]{wu2006recent}
\bibinfo{author}{WU, K.h.}, \bibinfo{author}{L{\"U}, X.d.}, \bibinfo{author}{Liu, Y.l.}, \bibinfo{year}{2006}.
\newblock \bibinfo{title}{Recent progress of pediatric cardiac surgery in china}.
\newblock \bibinfo{journal}{Chinese medical journal} \bibinfo{volume}{119}, \bibinfo{pages}{2005--2012}.
%Type = Article
\bibitem[{Wu et~al.(2014)Wu, Li, Xia, Ji, Liang, Ma, Li, Wu, Wang and Zhao}]{wu2014prevalence}
\bibinfo{author}{Wu, L.}, \bibinfo{author}{Li, B.}, \bibinfo{author}{Xia, J.}, \bibinfo{author}{Ji, C.}, \bibinfo{author}{Liang, Z.}, \bibinfo{author}{Ma, Y.}, \bibinfo{author}{Li, S.}, \bibinfo{author}{Wu, Y.}, \bibinfo{author}{Wang, Y.}, \bibinfo{author}{Zhao, Q.}, \bibinfo{year}{2014}.
\newblock \bibinfo{title}{Prevalence of congenital heart defect in guangdong province, 2008-2012}.
\newblock \bibinfo{journal}{BMC public health} \bibinfo{volume}{14}, \bibinfo{pages}{152}.
%Type = Article
\bibitem[{Yang et~al.(2018)Yang, Zhang, Xiang, Torr and Hospedales}]{yang2018learning}
\bibinfo{author}{Yang, F.S.Y.}, \bibinfo{author}{Zhang, L.}, \bibinfo{author}{Xiang, T.}, \bibinfo{author}{Torr, P.H.}, \bibinfo{author}{Hospedales, T.M.}, \bibinfo{year}{2018}.
\newblock \bibinfo{title}{Learning to compare: Relation network for few-shot learning} .
%Type = Inproceedings
\bibitem[{Yang et~al.(2017)Yang, Ren, Zhang, Chen, Wen, Li and Hua}]{yang2017neural}
\bibinfo{author}{Yang, J.}, \bibinfo{author}{Ren, P.}, \bibinfo{author}{Zhang, D.}, \bibinfo{author}{Chen, D.}, \bibinfo{author}{Wen, F.}, \bibinfo{author}{Li, H.}, \bibinfo{author}{Hua, G.}, \bibinfo{year}{2017}.
\newblock \bibinfo{title}{Neural aggregation network for video face recognition}, in: \bibinfo{booktitle}{IEEE CVPR}, pp. \bibinfo{pages}{4362--4371}.
%Type = Article
\bibitem[{Yang et~al.(2009)Yang, LI, L{\"U} and Liu}]{yang2009incidence}
\bibinfo{author}{Yang, X.y.}, \bibinfo{author}{LI, X.f.}, \bibinfo{author}{L{\"U}, X.d.}, \bibinfo{author}{Liu, Y.l.}, \bibinfo{year}{2009}.
\newblock \bibinfo{title}{Incidence of congenital heart disease in beijing, china}.
\newblock \bibinfo{journal}{Chinese medical journal} \bibinfo{volume}{122}, \bibinfo{pages}{1128--1132}.
%Type = Article
\bibitem[{Zhang et~al.(2018)Zhang, Gajjala, Agrawal, Tison, Hallock, Beussink-Nelson, Lassen, Fan, Aras, Jordan et~al.}]{zhang2018fully}
\bibinfo{author}{Zhang, J.}, \bibinfo{author}{Gajjala, S.}, \bibinfo{author}{Agrawal, P.}, \bibinfo{author}{Tison, G.H.}, \bibinfo{author}{Hallock, L.A.}, \bibinfo{author}{Beussink-Nelson, L.}, \bibinfo{author}{Lassen, M.H.}, \bibinfo{author}{Fan, E.}, \bibinfo{author}{Aras, M.A.}, \bibinfo{author}{Jordan, C.}, et~al., \bibinfo{year}{2018}.
\newblock \bibinfo{title}{Fully automated echocardiogram interpretation in clinical practice: feasibility and diagnostic accuracy}.
\newblock \bibinfo{journal}{Circulation} \bibinfo{volume}{138}, \bibinfo{pages}{1623--1635}.
%Type = Article
\bibitem[{Zhang et~al.(2015)Zhang, Zeng, Zhao and Lu}]{zhang2015diagnostic}
\bibinfo{author}{Zhang, Y.F.}, \bibinfo{author}{Zeng, X.L.}, \bibinfo{author}{Zhao, E.F.}, \bibinfo{author}{Lu, H.W.}, \bibinfo{year}{2015}.
\newblock \bibinfo{title}{Diagnostic value of fetal echocardiography for congenital heart disease: a systematic review and meta-analysis}.
\newblock \bibinfo{journal}{Medicine} \bibinfo{volume}{94}.
%Type = Article
\bibitem[{Zhao et~al.(2014)Zhao, Ma, Ge, Liu, Yan, Wu, Ye, Liang, Zhang, Gao et~al.}]{zhao2014pulse}
\bibinfo{author}{Zhao, Q.m.}, \bibinfo{author}{Ma, X.j.}, \bibinfo{author}{Ge, X.l.}, \bibinfo{author}{Liu, F.}, \bibinfo{author}{Yan, W.l.}, \bibinfo{author}{Wu, L.}, \bibinfo{author}{Ye, M.}, \bibinfo{author}{Liang, X.c.}, \bibinfo{author}{Zhang, J.}, \bibinfo{author}{Gao, Y.}, et~al., \bibinfo{year}{2014}.
\newblock \bibinfo{title}{Pulse oximetry with clinical assessment to screen for congenital heart disease in neonates in china: a prospective study}.
\newblock \bibinfo{journal}{The Lancet} \bibinfo{volume}{384}, \bibinfo{pages}{747--754}.
%Type = Article
\bibitem[{Zheng et~al.(2008)Zheng, Barbu, Georgescu, Scheuering and Comaniciu}]{zheng2008four}
\bibinfo{author}{Zheng, Y.}, \bibinfo{author}{Barbu, A.}, \bibinfo{author}{Georgescu, B.}, \bibinfo{author}{Scheuering, M.}, \bibinfo{author}{Comaniciu, D.}, \bibinfo{year}{2008}.
\newblock \bibinfo{title}{Four-chamber heart modeling and automatic segmentation for 3-d cardiac ct volumes using marginal space learning and steerable features}.
\newblock \bibinfo{journal}{IEEE transactions on medical imaging} \bibinfo{volume}{27}, \bibinfo{pages}{1668--1681}.
%Type = Article
\bibitem[{Zhou et~al.(2018)Zhou, Andonian and Torralba}]{zhou2017temporal}
\bibinfo{author}{Zhou, B.}, \bibinfo{author}{Andonian, A.}, \bibinfo{author}{Torralba, A.}, \bibinfo{year}{2018}.
\newblock \bibinfo{title}{Temporal relational reasoning in videos}.
\newblock \bibinfo{journal}{In ECCV} .
%Type = Inproceedings
\bibitem[{Zhou et~al.(2017)Zhou, Huang, Wang, Wang and Tan}]{zhou2017see}
\bibinfo{author}{Zhou, Z.}, \bibinfo{author}{Huang, Y.}, \bibinfo{author}{Wang, W.}, \bibinfo{author}{Wang, L.}, \bibinfo{author}{Tan, T.}, \bibinfo{year}{2017}.
\newblock \bibinfo{title}{See the forest for the trees: Joint spatial and temporal recurrent neural networks for video-based person re-identification}, in: \bibinfo{booktitle}{Computer Vision and Pattern Recognition (CVPR), 2017 IEEE Conference on}, \bibinfo{organization}{IEEE}. pp. \bibinfo{pages}{6776--6785}.
%Type = Article
\bibitem[{Zou et~al.(2019)Zou, Yu, Liu, Kumar and Wang}]{zou2019confidence}
\bibinfo{author}{Zou, Y.}, \bibinfo{author}{Yu, Z.}, \bibinfo{author}{Liu, X.}, \bibinfo{author}{Kumar, B.}, \bibinfo{author}{Wang, J.}, \bibinfo{year}{2019}.
\newblock \bibinfo{title}{Confidence regularized self-training}.
\newblock \bibinfo{journal}{ICCV} .

\end{thebibliography}

\end{document}